\documentclass[]{aa} 

\usepackage{graphicx,txfonts,float,rotating,color,url,lscape}

\newcommand{\less}{\raisebox{-1.1mm}{$\stackrel{<}{\sim}$}} 
 
\newcommand{\msol}{\mbox{M$_{\odot}$}}

\newcommand{\spi}{{\sigma}_{\pi}}
\newcommand{\G}{{\it Gaia}}
\newcommand{\Hp}{{\it Hipparcos}}

\begin{document}

\title{
The Cepheid period -- luminosity -- metallicity relation based on {\it Gaia} DR2 data
\thanks{
Table~1 is available in electronic form at the CDS via 
anonymous ftp to cdsarc.u-strasbg.fr (130.79.128.5) or via 
http://cdsweb.u-strasbg.fr/cgi-bin/qcat?J/A+A/. 
} 
}  
 
\author{ 
M.~A.~T.~Groenewegen
}

\institute{ 
Koninklijke Sterrenwacht van Belgi\"e, Ringlaan 3, B--1180 Brussels, Belgium \\ \email{martin.groenewegen@oma.be}
} 
 
\date{received: 23 May 2018, accepted: 23 July 2018} 
 
\offprints{Martin Groenewegen} 
 
 
\abstract
%
{} 
 {We use parallax data from the \it Gaia \rm second data release (GDR2), combined with parallax data based on \it Hipparcos \rm and \it HST \rm   data, 
to derive the period -- luminosity -- metallicity ($PLZ$) relation for Galactic classical cepheids (CCs) in the $V$, $K$, and Wesenheit $WVK$ bands.
}
 {An initial sample of 452 CCs are extracted from the literature with spectroscopically derived iron abundances.
Reddening values, classifications, pulsation periods, and mean $V$- and $K$-band magnitudes are taken from the literature.
Based on nine CCs with a goodness-of-fit (GOF) statistic smaller than 8 and with an accurate non-\it Gaia \rm parallax 
($\spi$ comparable to that in GDR2), a parallax zero-point offset of $-0.049 \pm 0.018$ mas is derived.
Selecting  a GOF statistic smaller than 8 removes about 40\% of the sample most 
likely related due to binarity. Excluding first overtone and multi-mode cepheids and applying some other criteria reduces the sample to about 200 stars.
}
   {
The derived $PL(Z)$ relations depend strongly on the parallax zero-point offset. 
The slope of the $PL$ relation is found to be different from the relations in the LMC 
at the  $3\sigma$ level. Fixing the slope to the value found in the LMC leads to a distance modulus (DM) to the LMC of order 18.7 mag, 
larger than the canonical distance. The canonical DM of around 18.5 mag would require a parallax zero-point offset of order $-0.1$ mas.
Given the strong correlation between zero point, 
period and metallicity dependence of the $PL$ relation, and the parallax zero-point offset there is no evidence for a metallicity term in the $PLZ$ relation.
}
   {The GDR2 release does not allow us to improve on the current distance scale based on CCs. 
The value of and the uncertainty on the parallax zero-point offset leads to uncertainties of order 0.15 mag on the distance scale.
The parallax zero-point offset will need to be known at a level of 3~$\mu$as or better to have a 0.01 mag or smaller effect 
on the zero point of the $PL$ relation and the DM to the LMC.
}


\maketitle

\section{Introduction} 

Classical Cepheids (CCs) are considered  important standard candles because they are
bright and thus the link between the distance scale in the nearby
universe and that further out via those galaxies that contain both Cepheids and SNIa 
(e.g. \citealt{Riess16} for a recent overview on how to get the Hubble constant to 2.4\% precision).

Distances to local CCs may be obtained in several ways, for example through direct determination of the parallax (see below) or
main-sequence fitting for Cepheids in clusters (e.g. \citealt{Feast1999, Turner10} for overviews).
In addition, distances to CCs can be obtained from the Baade--Wesselink (BW) method. This method relies on the availability of
surface-brightness (SB) relations to link variations in colour to variations in angular diameters and an understanding of the 
projection ($p$-) factor, which links radial velocity to pulsational velocity variations. This method is interesting for more 
distant cepheids where an accurate direct parallax determination is not possible. The most recent works for 70--120 Galactic 
and about 40 Magellanic Cloud cepheids are by \cite{Storm11a,Storm11b} and \cite{Gr2013}. 
These papers also investigated the possible metallicity dependence of the period--luminosity ($PL$) relation, 
which is one of the remaining possible sources of systematic  uncertainties in the application of the $PL$ relation to
the distance scale. Although the effect is deemed to be subdominant (0.5\% on a total uncertainty of 2.4\% in the 
determination of the Hubble constant, as stated by \citealt{Riess16}), estimates in the literature for its actual value 
and error estimate vary considerably and seem to depend on wavelength (see \citealt{Storm11b} and \citealt{Gr2013} for references) 
and a closer investigation is certainly in order in the general framework of `precision cosmology' and a 1\% accurate 
Hubble constant.

As accurate direct distances to a sizeable number of Galactic Cepheids were unavalaible pre-\G\ the BW method was 
the only way to investigate this.
Both papers agree that the metallicity dependence in the $K$  band is statistically insignificant with the data they had. 
Storm et al. found a 2$\sigma$ effect in the classical Wesenheit relation based on $V, I$ [$W(VI)= V - 2.55\;(V-I)$)], while
Groenewegen found a 2$\sigma$ effect in the $V$ band.

These types of questions can be addressed directly  when accurate parallaxes are available for a significant sample of Galactic CCs.
The Gaia second data release (GDR2, \citealt{GDR2Sum}) extends GDR1 \citep{GC2016a,GC2016b}.
The Gaia parallaxes on CCs extend earlier work based  on \Hp\ parallaxes \citep{ESA1997,vanL07,vanL07NR,vanL08}, 
and parallel work using the {\it Fine Guidance Sensor} \citep{Benedict07} and the {\it Wide Field Camera 3} \citep{Riess14,Casertano16,Riess18} 
on board the {\it Hubble Space Telescope} (HST) for about 20 CCs.

In this paper we aim to investigate the $PL$ relation and its possible metallicity dependence based on a sample of Galactic cepheids 
available in the {\it Gaia} DR2.
The paper is structured as follows. In Section~\ref{S-PRE} the collection of photometric, reddening, metallicity, and other data from the sample is described.
Section~\ref{S-GDR2} describes the data taken from GDR2, and compares periods and classifications from the literature with those 
provided in GDR2.
The method used in the analysis in described in Section~\ref{S-Ana}, and tested with simulations in Section~\ref{S-Sim}.
Section~\ref{S-Res} presents the results which are summarised and discussed in Section~\ref{S-Dis}.

\section{Pre-{\it Gaia} DR2 preparation} 
\label{S-PRE}

The preparation for this paper started with the collection from the literature of all CCs with 
individually determined accurate iron abundances from high-resolution spectroscopy (see below for detailed references). 
This resulted in a sample of 452 stars, and the data described below are listed in Table~\ref{Tab:Targets}.

To perform the study on the metallicity dependence of the $PL$ relation, the following data are required: the classification 
(Cepheids can be fundamental mode (FU) pulsators,  first overtone (FO) or second overtone (SO) pulsators, or double-mode (DM) pulsators); 
the pulsation period; magnitudes (in this paper we concentrate on $V$, and the near-IR magnitudes $JHK$); 
the reddening $E(B-V)$ to de-redden the photometry; 
the metallicity (synonymous here with the iron abundance [Fe/H]); and the parallax.

Mean $V$ magnitudes are taken mainly from \citet{Melnik15}. 
This reference provides standard Johnson $V$ magnitudes for a sample of 674 cepheids, the latest extension of 
the collection of optical photometry following \citet{Berdnikov2000}.
Only 28 of the stars are not listed there. 
The footnote to Table~\ref{Tab:Targets} contains details on the photometry that was used. 
In some cases mean magnitudes were derived  by fitting Fourier series to time series data 
using the {\sc Period04} software \citep{Period04}. From a comparison of the mean magnitudes quoted in different sources, an 
error in the mean magnitude of 0.008 mag is adopted.

The near-IR (NIR) photometry is more heterogeneous as it comes from a variety of sources, using different photometric systems and 
ranges from intensity-mean magnitudes from well-sampled light curves to single-epoch photometry in some cases.
In order of preference, mean magnitudes are taken from 
\citet{MP11}, converted to the 2MASS system based on the transformation equations in their Table~1); 
SAAO-based photometry (mainly \citealt{Laney1992}, and Laney (priv. comm.), as quoted in \citet{Genovali2014} and \citealt{Feast2008}), 
converted to the 2MASS system based on the transformation equations in \citet{Koen2007}; and
CIT-based photometry from \citet{Welch1984} and \citet{Barnes1997}, converted to the 2MASS system based on the 
transformation equations in \citet{MP11}.
For the remaining sources the median was taken of the available single-epoch data available in
\citet{McGonegal83}, \citet{Welch1984}, \citet{Schechter1992}, DENIS ($JK$ data transformed to 
the 2MASS system using \citealt{Carpenter2001}), and 2MASS.

For the data by \citet{MP11} and the SAAO-based data an error in the mean magnitude of 0.008 mag was assumed.
For the mean magnitudes from CIT-based data an error of 0.01 mag was assumed as they generally appear to be of slightly lower quality. 
For the photometry based on median filtering of multiple observations an error of 0.025 mag was assumed.
If only a single-epoch 2MASS observation was available a typical error of 0.025 mag was assumed, 
unless the quality flag was not AAA, in which case a typical error of 0.25 mag was assumed.

A special case is Polaris. The 2MASS magnitude is highly uncertain ($K= 0.456 \pm 0.248$ mag). The {\it COBE-DIRBE} flux at 
1.25 and 2.2 $\mu$m was taken \citep{COBED04} and converted to magnitudes using the 2MASS zero points (ZPs). 
Including error bars in the flux and in the ZPs we arrives at $J= 0.941 \pm 0.031$ and $K= 0.652 \pm 0.028$ mag. 
As Polaris is hardly variable, this is essentially an estimate of the mean intensity.
This value is consistent with the older photometry by \citet{Gehrz1974}. 
Taking the ZP of that system \citep{GH1974}, and converting the flux back to
a 2MASS magnitude we arrive at an estimate $K= 0.64 \pm 0.10$ mag.

Reddening values, $E(B-V)$, and the error therein are primarily taken from the compilation in 
\citet{Fernie1995}\footnote{ \url {http://www.astro.utoronto.ca/DDO/research/cepheids/table_colourexcess.html} } with a scaling factor as indicated below.
Only about 50 stars are not listed there.

Tammann et al. (2003) suggested  scaling the values in \citet{Fernie1995} by a factor of 0.951 to have consistency between the values listed there and those 
derived from a period-colour relation.
\citet{Fouque07} also discussed reddening and adopted the reddening from \citet{LanCal07} based on $BVI$ photometry, which is also adopted here.
\citet{Fouque07} find a scaling factor of 0.952 $\pm$ 0.010 with respect to the reddenings listed in \citet{Fernie1995}.
The stars in \citet{Fouque07} were compared to those in \citet{LanCal07}. Thirty-nine are in overlap, of which 27 have $E(B-V) > 0.2$ mag.
The ratio of the reddenings lies between 0.91--1.18 and both the median and mean ratio are 1.00 with a dispersion of 0.03.
Comparing \citet{Fouque07} to \citet{Fernie1995} there are 127 stars with $E(B-V) > 0.2$  mag in overlap, with a range in ratios of 0.80--1.42  with 
mean and median of 0.93--0.94 and dispersion 0.05.
Similarly, comparing \citet{Tammann03} to \citet{Fernie1995} there are 184 stars with $E(B-V) > 0.2$  mag in overlap, with a range in ratios of 0.78--1.16  with 
mean and median of 0.94--0.95 and dispersion 0.05.

In order of preference, reddenings were taken from \citet{Fernie1995} scaled by a factor of 0.94; 
\citet{Acharova2012} without scaling,  \citet{LL11} scaled by 0.99, \citet{CC87} scaled by 0.987), \citet{Kashuba16} scaled by 0.94,
\citet{MarAnd15} scaled by 0.97, and \citet{Sziladi07} scaled by 0.92. 
For eight stars no reddening appears to have been published, and these were estimated
from several 3D reddening models \citep{Marshall06, Drimmel2003, Arenou1992} 
using the parallax from GDR2 (see \citealt{Gr2008} for details).

The error in $E(B-V)$ is taken from \citet{Fernie1995} or from the spread among the different 3D reddening estimates. 
Otherwise it is assumed to be $0.1 \, E(B-V)$.
The extinction in the visual is assumed to be $A_{V}= 3.1 E(B-V)$, and extinction ratios 
$A_{J}/A_{V}$, $A_{H}/A_{V}$, and $A_{K}/A_{V}$ of 0.276, 0.176, and 0.118, respectively, have been adopted.

The iron abundances are taken from several sources and put on the same scale. The main source is the compilation by \citet{Genovali2014}, which has data for 434 stars when 
combined with \citet{Genovali2015}.
They compared iron abundances from different literature sources and re-scaled all data to a uniform scale.
Column~12 in Table~\ref{Tab:Targets} lists the iron abundance, if available, from  \citet{Genovali2014} or 
from the follow-up work in \citet{Genovali2015}.

Another large compilation is that in \citet{Ngeow12}, which has 329 stars in common with \citet{Genovali2014,Genovali2015}. 
The average difference (in the sense Ngeow - Genovali et al.) = +0.02 $\pm$ 0.06 dex.  Ngeow works on the system by Luck, Lambert, and coworkers, 
and therefore a direct comparison is made to the iron abundances in \citet{LL11}. There are 318 stars in common with 
\citet{Genovali2014,Genovali2015}. The average difference (in the sense Luck \& Lambert - Genovali et al.) = +0.03 $\pm$ 0.05 dex.
Column~14 in Table~\ref{Tab:Targets} lists the iron abundance, if available, from \citet{Ngeow12} or \citet{LL11} without further correction.

Some other catalogues were also considered. The large compilation by \citet{Acharova2012} has 277 stars in common with \citet{Genovali2014,Genovali2015}.
The average difference (in the sense Acharova et al. - Genovali et al.) is  $-0.055$ $\pm$ 0.08 dex.
\citet{Sziladi07} has 14 stars in common with \citet{Genovali2014}.
The average difference (in the sense Szil\'adi et al. - Genovali et al.) is  $-0.032$ $\pm$ 0.07 dex.
\citet{MarAnd15} has 22 stars in common with \citet{Genovali2014}.
The average difference (in the sense Martin et al. - Genovali et al.) is  $-0.03$ $\pm$ 0.07 dex.
Column~16 in Table~\ref{Tab:Targets} lists the iron abundance, if available, from these three references, with offsets applied.
In the analysis below the value in Col.~12 is preferred over that in Col.~14, which is preferred over that in Col.~16.
Based on the comparison between datasets and the scatter between different measurements, an error of 0.08 dex in [Fe/H] is assumed.

Regarding the variability type and pulsation period the Variable Star indeX catalogue (VSX, \citealt{Watson06}) was the main source of information, but 
other sources were also consulted \citep{Berdnikov2000, Klagyivik2009, LL11, Ngeow12, Genovali2014, Melnik15}.
Periods agree typically to a high degree, of order $4 \cdot 10^{-4} P$ or better. Pulsation types are sometimes less certain. 
This can be related to the FU or FO classification, or even the classification as CC. 
The star V473 Lyr is assumed to be a SO CC \citep{Molnar17}.

The stars BC Aql, TX Del, AU Peg, and SU Sct are classified as (likely) Type-II Cepheids (T2Cs). 
 QQ Per is also marked as an uncertain CC (indicated by the `?') and has been classified as a T2C as well.
Two stars have a very different classification. The star EK Del is classified as a possible  Above the Horizontal Branch (AHB) star. 
Its metallicity of  [Fe/H]= $-1.57$ dex is by far the lowest among the 452 objects and seems more closely related to that of RR Lyrae.
The object V1359 Aql is classified as `ROT', i.e. a spotted star whose variability is due to rotation, with a period of 96.3 days.
These stars were kept in the sample, anticipating that the GDR2 would also contain classifications for many variables (see next section and Table~\ref{Tab:Types}).

The total number of stars that is potentially  used for the analysis of the $PLZ$ relation is 426; the starting sample of 
452 listed in Table~\ref{Tab:Targets}, minus 2 targets not listed in GDR2 (see next section), minus 6 stars almost certainly not CCs, 
and minus 18 stars that are SO or DM Cepheids that were also a priori excluded.

\section{{\it Gaia} DR2 data} 
\label{S-GDR2}

The data was obtained by querying the various tables through VizieR.
The list of objects was cross-matched with the {\tt gaiadr2.gaia\_source} table using a radius of 1.2\arcsec.
The largest differences were for CE Cas A (at 0.9\arcsec) and Polaris (at 0.6\arcsec). The other sources were matched to
within 0.3\arcsec\ or better. Two sources, V340 Nor and IY Cep, were not found (even when a larger search radius was used),
and they appear to be missing from GDR2.

From the source table the following parameters were retrieved:

\begin{itemize}

\item The unique source identifier {\tt source\_id} for querying other tables (see below);

\item The {\tt parallax} ($\pi$) and {\tt parallax\_error} ($\spi$) (both in mas); 

\item Parameters describing the quality of the astrometric fit, in particular,

1) the  goodness-of-fit  (GOF) statistic, {\tt astrometric\_gof\_al}, of the astrometric solution for the source in 
the along-scan direction. For good fits it should approximately follow a normal distribution with zero mean value
and unit standard deviation; 

2) the {\tt astrometric\_excess\_noise}, $\epsilon_i$, which 
is quadratically added to the assumed observational noise in each observation in order to statistically match the residuals in the astrometric solution
and
the {\tt astrometric\_excess\_noise\_sig}, $D$, the significance of $\epsilon_i$, where 
`A value $D > 2$ indicates that the given $\epsilon_i$ is probably significant'\footnote{See the Gaia Data Release 2 documentation available 
at \url{https://gea.esac.esa.int/archive/documentation/GDR2/pdf/GaiaDR2_documentation_1.0.pdf} for a description of these parameters.}.

\medskip
We note that the catalogued values for these parameters have {not} been corrected for the `DOF bug', as discussed in Appendix~A in \citet{Lindegren18}.

\medskip
The GOF parameter was also given in the \Hp\ data releases, but was not listed in GDR1. 
The excess noise parameter and its significance  were parameters introduced in GDR1.

\end{itemize}

\noindent
GDR2 provides additional information, in particular regarding variability (see \citealt{GDR2Var}). Two types of classification and analysis are available. 
The first is based on at least two transits, the {\tt nTransits:2+} classifier, which gives a {\tt best\_class\_name,} and a {\tt best\_class\_score}, 
a number between 0 and 1, indicating the confidence of the classification.
More useful information is available when more transit data is available and the objects are passed through  Specific Objects Studies (SOS) modules.
In particular a total of 9575 objects have been classified as a Cepheid by the SOS module on Cepheids and RR Lyrae \citep{Clementini18}. 
Based on the  {\tt source\_id} the {\tt vari\_cepheid} table was queried to return the following:

\begin{itemize}

\item The {\tt type\_best\_classification}, which can be DCEP, T2CEP, and ACEP  respectively for  Classical Cepheids, Type-II Cepheids, and Anomalous Cepheids;

\item The {\tt mode\_best\_classification}, which can be FUNDAMENTAL, FIRST OVERTONE, or MULTI;

\item The pulsation period with error (In the case of a MULTI classification two periods are given);

\item The metallicity of the star derived from the Fourier parameters of the light curve, and its error.

\end{itemize}

Of the sample 257 are classified as FU pulsators, 43 as FO, 6 as multi-mode, and 8 as T2C by the SOS module.
The {\tt nTransits:2+} classifier lists 5 Anomalous Cepheids, 50 T2C, 300 CCs, and 7 Mira/Semi-regular pulsators.

\begin{figure}

\begin{minipage}{0.24\textwidth}
\resizebox{\hsize}{!}{\includegraphics{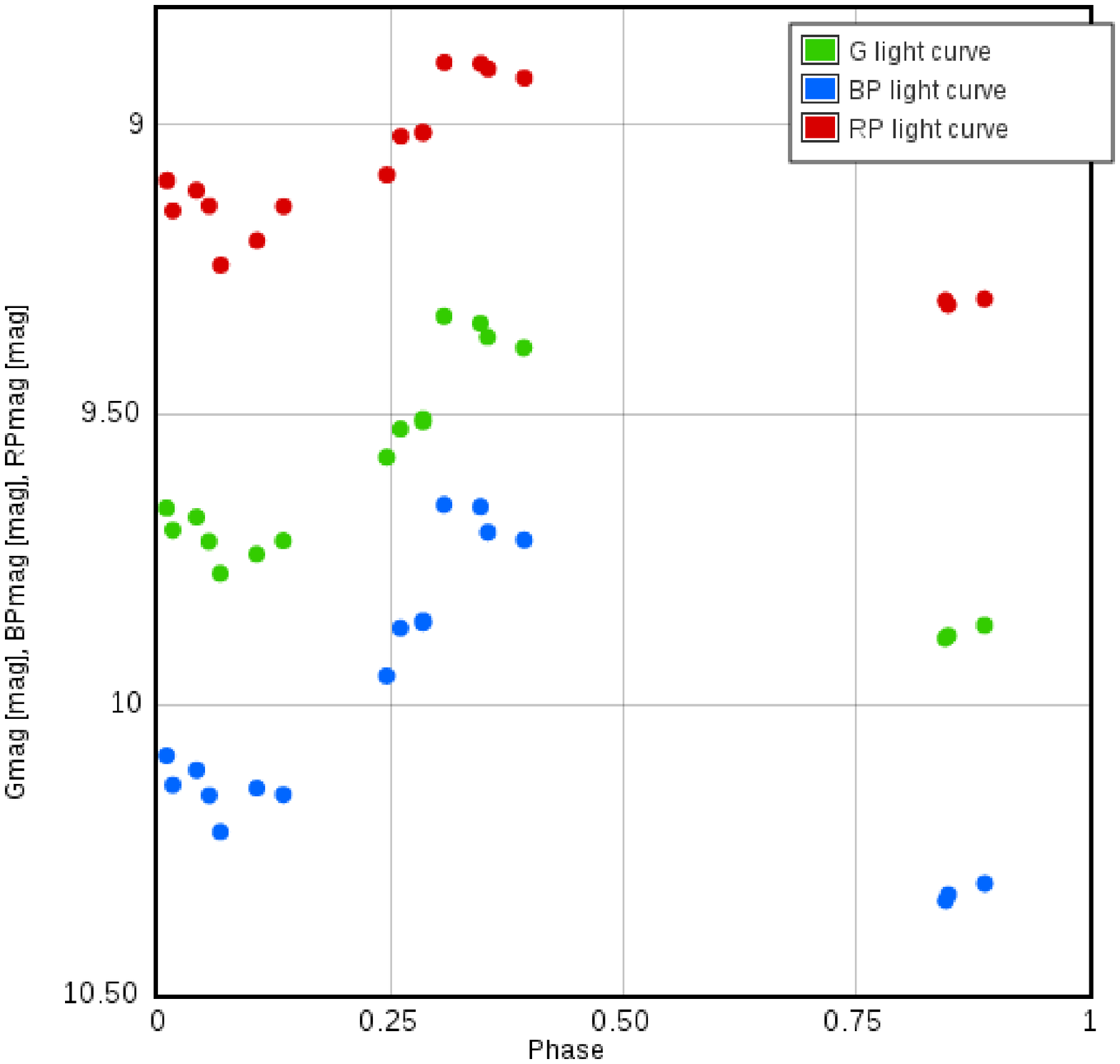}}
\end{minipage}
\begin{minipage}{0.24\textwidth}
\resizebox{\hsize}{!}{\includegraphics{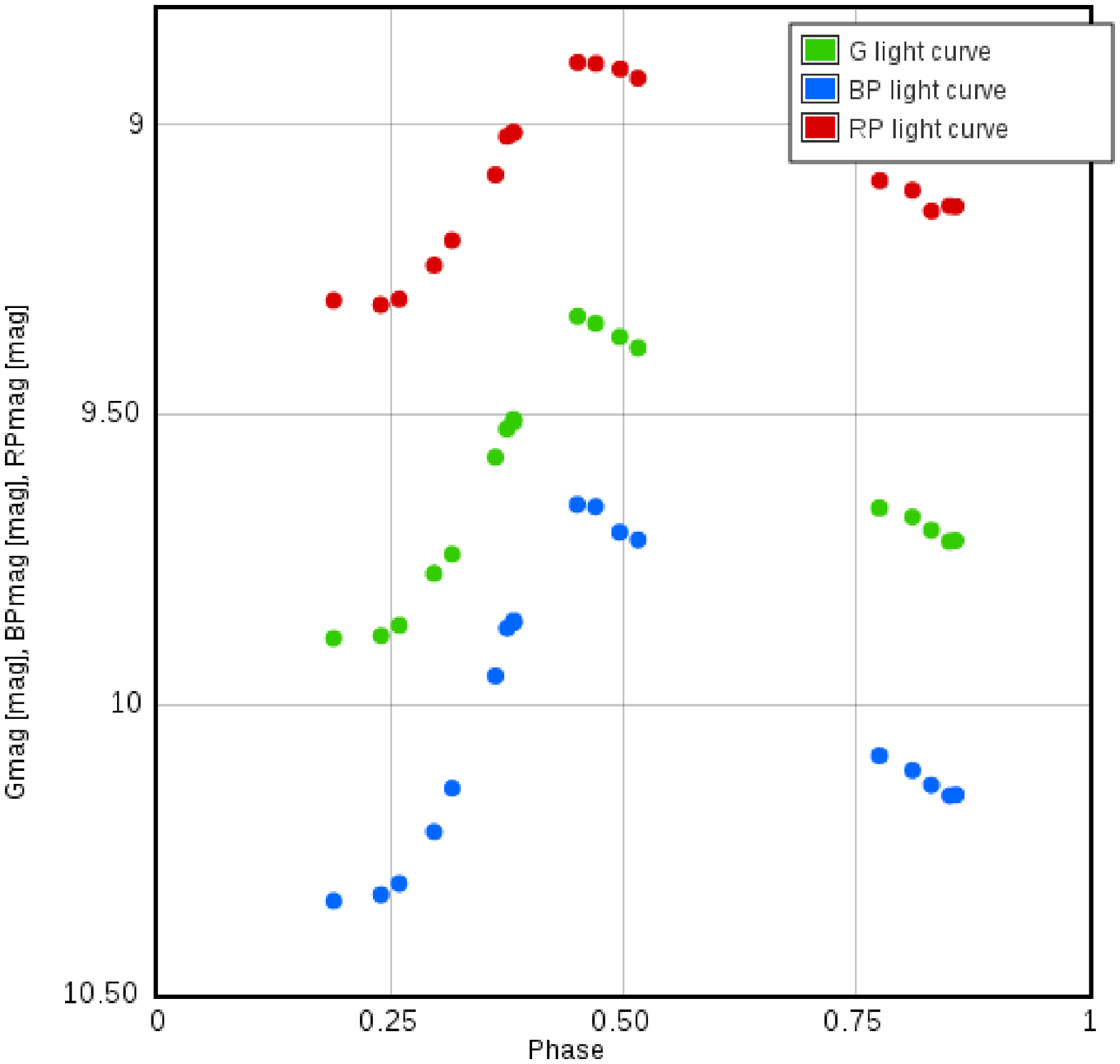}}
\end{minipage}

\caption{Example of the phased light curve for AD Gem in the {\it Gaia} $G$, $Bp$, and $Rp$ bands fitted 
with the period as given by the SOS (1.895 days, left) and the correct period (3.788 days,  right). 
The plots were made using the tools available through VizieR.
}
\label{Fig:Periods}
\end{figure}

Based on the data in GDR2 some checks were performed against the data that was prepared pre-{\it Gaia}.
%
%
For 279 stars the periods derived by the SOS module compare well to the value in the literature (in Table~\ref{Tab:Targets}), 
with a relative precision better than 0.5\% and a median value $\mid \Delta P\mid/P = 1.3 \cdot 10^{-4}$.
Then there is a jump, and for 35 stars the periods are significantly different with $\mid \Delta P\mid/P > 0.06$.
The most extreme example is V500 Sco with a true period of 9.3 and a derived period of 175.6 days.
Other periods differ by a near integer number, for example T Vul (true period of 4.435 and  a derived period of 2.217 days) or
XX Car (true period of 15.716 and a derived period of 31.449 days).
All 35 cases were inspected by folding the Gaia light curve  with the period in the literature;  the literature period
always fits the Gaia light curve better (see Fig.~\ref{Fig:Periods} for an example).

In the present sample about 11\% of the Cepheids have been assigned an incorrect period. As the paper describing the 
SOS Cepheid and RR Lyrae module used for GDR2 \citep{Clementini18} does not contain any information about 
the period validation for Cepheids it is unclear how representative this fraction is.

Of interest is also the classification of the objects, and their pulsation mode.
Of the 314 stars in the sample analysed by the SOS, 271 classifications agree with the value in the literature. 
The other 43 cases are listed in Table~\ref{Tab:Types}. The most common difference is between the FU and FO pulsations.
As noted above some periods are incorrect, and this is indicated as it might have influenced the classification as well.

\setcounter{table}{1}
\begin{table}
\setlength{\tabcolsep}{1.4mm}
 \caption{Comparison of pulsation types and modes.}
\centering
\small
  \begin{tabular}{rrrrl}
  \hline
Name   &   Literature  &  GDR2         &  GDR2         &  Remarks  \\ 
       &   (Tab.~1)    &  SOS          &  {\tt nTran}  &    \\ 
  \hline

BG Cru &    DCEPS       &  DCEP-FU  &  CEP         & \\
CI Per &    DCEP?       &  DCEP-FO  &  T2CEP       & \\
CR Cep &    DCEP?       &  DCEP-FU  &  CEP         & \\
CY Aur &    DCEP        &  DCEP-FO  &  CEP         & Period incorrect \\
DK Vel &    DCEP        &  DCEP-FO  &  CEP         & \\
FM Aql &    DCEP        &  DCEP-FO  &  CEP         & Period incorrect \\
FO Cas &    DCEP        &  T2CEP    &  T2CEP       & \\
GH Car &    DCEPS       &  DCEP-FU  &  CEP         & \\
IT Car &    DCEPS       &  DCEP-FU  &  CEP         & \\
MY Pup &    DCEPS       &  DCEP-FU  &  CEP         & \\
NT Pup &    DCEP        &  T2CEP    &  T2CEP       & \\
RS Ori &    DCEP        &  DCEP-FO  &  CEP         & Period incorrect   \\
RW Cam &    DCEP        &  T2CEP    &  CEP         & \\
TT Aql &    DCEP        &  DCEP-FO  &  CEP         & Period incorrect \\
TU Cas &    DCEP(B)     &  DCEP-FU  &  CEP         & \\
TX Del &    CWB:        &  T2CEP    &  CEP         & \\
V1334 Cyg &   DCEPS     &  DCEP-FU  &  CEP         & \\
V350 Sgr &    DCEP      &  DCEP-FO  &  CEP         & Period incorrect  \\
V378 Cen &    DCEPS     &  DCEP-FU  &  CEP         & \\
V482 Sco &    DCEP      &  DCEP-FO  &  CEP         & Period incorrect \\
V500 Sco &    DCEP      &  T2CEP    &  CEP         & Period incorrect \\
V636 Cas &    DCEPS     &  DCEP-FU  &  CEP         & \\
V659 Cen &    DCEPS     &  DCEP-FU  &  CEP         & \\
V924 Cyg &    DCEPS     &  DCEP-FU  &  CEP         & \\
 X Lac &    DCEPS       &  DCEP-FU  &  CEP         & \\
 Y Oph &    DCEP?       &  DCEP-FU  &  CEP         & \\
 Y Car &    DCEP(B)     &  DCEP-FO  &  CEP         & \\
GZ Car &    DCEP(B)     &  DCEP-FU  &  CEP         & \\
BK Cen &    DCEP(B)     &  DCEP-FU  &  CEP         & \\
V458 Sct &  DCEP(B)     &  DCEP-FU  &  CEP         & Period incorrect \\
 U TrA &    DCEP(B)     &  DCEP-FU  &  CEP         & \\
V493 Aql &  DCEP        &  MULTI    &  CEP         & Period incorrect \\ 
V526 Aql &  DCEP        &  T2CEP    &  T2CEP       & Period incorrect \\
CO Aur &    DCEPS(B)    &  DCEP-FO  &  CEP         & \\
AC Cam &    DCEP        &  MULTI    &  -  & \\
FW Cas &    DCEP        &  MULTI    &  -  & \\
HK Cas &    DCEP        &  DCEP-FO  &  ACEP        & \\
EK Del &    AHB1:       &  DCEP-FU  &  ACEP        & \\
FQ Lac &    CEP:?       &  DCEP-FU  &  T2CEP       & \\
BE Mon &    DCEP        &  DCEP-FO  &  CEP         & \\
QQ Per &    CEP?        &  T2CEP    &  T2CEP       & \\
CR Ser &    DCEP        &  DCEP-FO  &  CEP         & Period incorrect \\
V1954 Sgr & DCEP        &  DCEP-FO  &  CEP         & Period incorrect \\

\hline
\end{tabular}

\tablefoot{
Pulsation type and mode from the literature (Col.~2, as listed in Table~\ref{Tab:Targets}) and from the SOS module (Col.~3).
In Col.~3 the classification from the {\tt nTransits:2+} classification scheme.
}

\label{Tab:Types}
\end{table}

\begin{figure}

\centering
\begin{minipage}{0.39\textwidth}
\resizebox{\hsize}{!}{\includegraphics{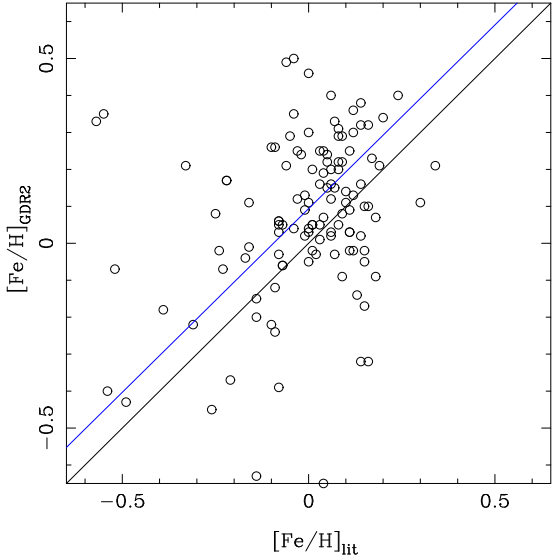}}
\end{minipage}

\caption{Comparison of the [Fe/H] abundance from the literature to that of the SOS. The black line is the one-to-one relation. 
The blue line is the bi-sector fit to the data.
}
\label{Fig:Iron}
\end{figure}

Interestingly, the SOS module also provides an iron abundance estimate based on the shape of the $G$-band light curve 
for 120 objects in the sample.
Figure~\ref{Fig:Iron} compares the adopted [Fe/H] abundance from the literature listed in Table~\ref{Tab:Targets} with the value 
provided in GDR2. There is a correlation, but with a lot of scatter. A bi-sector fit gives a slope of 0.99 and a zero point of 0.09.
The scatter around this relation is 0.26 dex, comparable to the quoted uncertainty of 0.22--0.24 dex (which includes a systematic error of 0.2 dex; see \citealt{Clementini18}).
This justifies the choice of considering only Cepheids with spectroscopic abundance determinations.

Table~\ref{Tab:Hipp} lists the CCs with accurate external parallaxes (i.e. non-\G, non-\Hp) and compares them to
\G\ DR2, \G\ DR1, and \Hp\ parallaxes. The external parallaxes are mostly based on {\it HST} \citep{Benedict07,Riess14,Casertano16,Riess18}. 
Out of interest, the parallax of Polaris~B is also listed, with the results from GDR2 and the parallax as recently determined by \citet{Bond18}. 
Polaris~B  is thought to be physically related to $\alpha$ UMi (see \citealt{Anderson18} and discussion therein). 
Another parallax estimate of 10.1 $\pm$ 0.2 mas exists for Polaris~A \citep{Turner13} based on the claimed membership of a cluster, but this
result has been disputed \citep{vanL13}. The photometric parallax of Polaris predicted by the $PL$ relation derived in the present paper is 
discussed in Appendix~\ref{AppPol}.

\begin{figure}
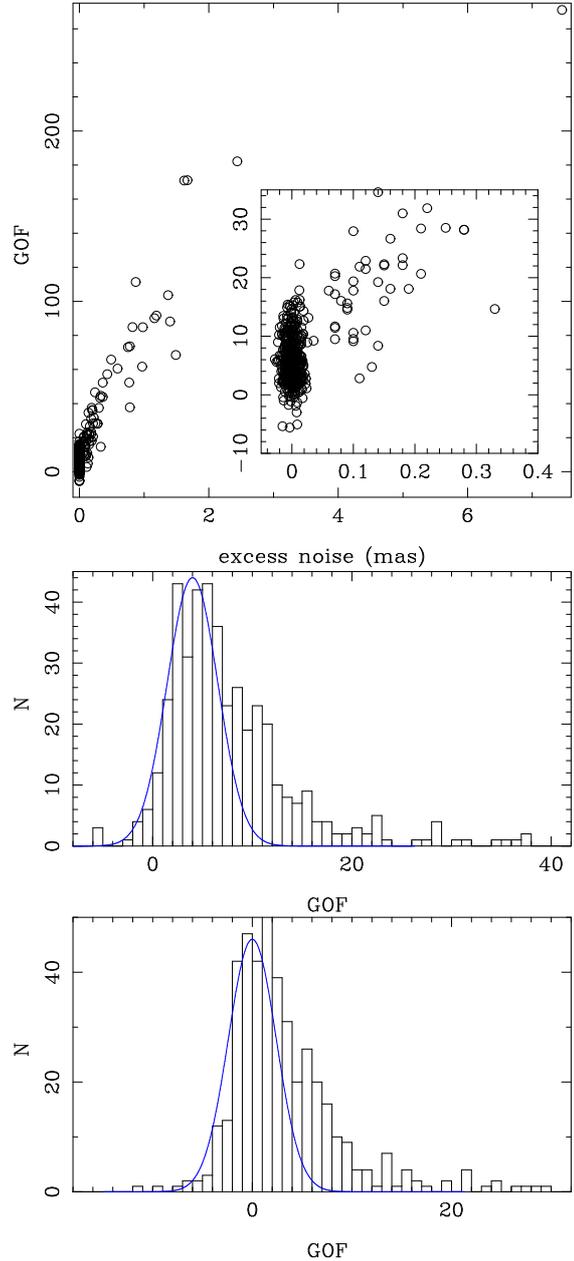


\begin{minipage}{0.4\textwidth}
\resizebox{\hsize}{!}{\includegraphics{GoFExNoComp.ps}}
\end{minipage}

\begin{minipage}{0.4\textwidth}
\resizebox{\hsize}{!}{\includegraphics{GoFHist.ps}}
\end{minipage}
\begin{minipage}{0.4\textwidth}
\resizebox{\hsize}{!}{\includegraphics{GoFHistCorr.ps}}
\end{minipage}

\caption{Goodness-of-fit (GOF) parameter plotted against the excess noise, with the inset showing a zoomed-in version.
At the extreme end is Polaris with $\epsilon_i= 7.5$ and GOF= 270 (top panel).
For clarity a randomly Gaussian distributed number is added to the excess noise when it is below 0.02 mas in the inset.
In the middle panel the histogram of the GOF parameter is shown with a Gaussian distribution (with mean 4.0 and $\sigma =1.8$) overplotted in blue.
In the bottom panel the GOF statistic has been recomputed by multiplying the  {\tt astrometric\_chi2\_al} statistic by 0.7 for all sources
to force a peak in the histogram at about zero. The $\sigma$ of the Gaussian is 1.7. 
The abscissa has been shifted by ten units compared to the plot in the middle panel.
}
\label{Fig:GOF}
\end{figure}

What is remarkable in Table~\ref{Tab:Hipp} is that many of the well-known Cepheids have very poor solutions with very large GOF and excess noise values.
There almost seems to be a dichotomy with Cepheids with zero excess noise to have a GOF  smaller than 8.
Figure~\ref{Fig:GOF} shows the relation between these two parameters for the entire sample, and the distribution over the GOF parameter.
We note that  this distribution is not a Gaussian with mean zero and unit variance, but it is due to the fact that this parameter
was not updated after discovery of the DOF bug \citep{Lindegren18}.

The bottom panel in Figure~\ref{Fig:GOF} shows the distribution over the GOF parameter when it is recomputed multiplying the {\tt astrometric\_chi2\_al} 
statistic by a constant for all sources and to force a peak in the histogram near zero. The factor used is 0.7, which is roughly consistent 
with the information provided in Appendix~A in \citet{Lindegren18}.

All nine stars with a GOF smaller than 8 have an accurate external parallax ($\spi$ comparable to that in GDR2).
The weighted mean difference (in the sense GDR2-external parallax) is $-0.049 \pm 0.018$ mas.
It is tempting to relate this to the parallax zero-point offset observed for QSO ($-0.029$ mas, \citealt{Lindegren18}), 
based on RGB stars from {\it Kepler} and {\it APOGEE} data  (about $-0.053$ mas, \citealt{Zinn18}), 
eclipsing binaries ($-0.082 \pm 0.033$ mas, \citealt{Stassun18}),
a sample of 50 CCs ($-0.046 \pm 0.013$ mas, \citealt{RiessGDR2}), 
RR Lyrae stars ($\sim -0.056$ mas, \citealt{Muraveva18}),
and the value of $-0.0319 \pm 0.0008$ mas mentioned for Cepheids in the GDR2 catalogue validation paper \citep{Arenou18}.

It should be noted that the assumption of a constant parallax zero-point offset is an oversimplification.
\citet{Lindegren18} already show that there are correlations with position on the sky, and trends with magnitude and colour
(their Figs.~7, 12, 13).

Binarity is common among Cepheids and has not been discussed so far. Binarity is not considered in solving for the astrometric parameters in GDR2. 
If binarity has an effect it would express itself in a poor fit when only solving for position, proper motion, and parallax.
As noted in Appendix~A in \citet{Lindegren18} the statistical quantities {\tt astrometric\_chi2\_al} and the GOF statistic 
have not been corrected for the DOF bug.  The GOF statistic is expected to follow a normal distribution around zero with unit variance, but
for the current sample it roughly follows a normal distribution which peaks near 4 and with a clear excess of stars with a GOF $>8$.
The models in Appendix~\ref{AppIni} show that selecting on GOF $<8$ removes 40\% of the sample. 
Many of those are known binaries.
\citet{RiessGDR2}  identify three outliers in their sample of 50 CCs, based on the location in a simple $\spi$ versus $\pi$ plot: SV Per, RW Cam, and RY Vel.
The GOF statistic of these objects is 84, 85, and 38, respectively, and  SV Per and RW Cam show close companions in their {\it HST} images.
Other known binaries have large GOF statistics and are therefore excluded from the analysis: V1334 Cyg (GOF=37) from  \cite{Evans2000} and \cite{Gallenne13}; 
AX Cir (GOF=387), KN Cen (GOF=8.5), SY Nor (GOF=13), AW Per (GOF=8.6), and SV Per (GOF=85) from \citet{Evans94}, 
and
R Cru (GOF=92) and S Mus (GOF=60) from \citet{Evans16}.
It appears that selecting a GOF $<8$ is an effective way of removing binaries from GDR2 data, and suggests that the results are not 
systematically influenced by binarity in the remaining sample.

\begin{sidewaystable*}
\setlength{\tabcolsep}{1.2mm}
 \caption{Comparison of {\it Gaia} data to {\it Hipparcos} and {\it HST} parallaxes.}
\centering
\small
  \begin{tabular}{rrrrrrrrrrrrrrrrrrrrrcccccccccc}
  \hline\hline
  Name       &  $\pi \pm \spi$  &  GOF & $\epsilon _i$  & D  &   $\pi \pm \spi$   &  $\epsilon _i$  & R & $\pi \pm \spi$  &  GOF & R & $\pi \pm \spi$  &  GOF  & R &  $\pi \pm \spi$  & R &  $\pi \pm \spi$  & R &  $\pi \pm \spi$ & R \\ 
             &     (mas)        &      &   (mas)         &   &       (mas)        &    (mas)        &   &     (mas)       &      &   &       (mas)     &       &   &      (mas)       &   &    (mas)         &   &    (mas)  &    \\ 
\hline
$\alpha$ UMi &  $\ldots$        & 271.04 &  7.46 & 15200 &  $\ldots$ & $\ldots$      &  - &   7.56 $\pm$  0.48 &  1.22 & 3 &   7.54 $\pm$  0.11 &  1.08 & 4 &  7.72 $\pm$  0.12 & 5 &  $\ldots$  &  -  & 7.620 $\pm$ 0.080 & 3,4,5 \\
Polaris~B    &   7.292 $\pm$  0.028 &  12.22 &  0.00 &     0 &  $\ldots$    &  $\ldots$     &  -  &  $\ldots$      &  $\ldots$     & - &  $\ldots$  & $\ldots$  & -  &  $\ldots$ & - &  6.26  $\pm$ 0.24 & 10 & $\ldots$   &  -            \\
$\beta$ Dor  &   3.112 $\pm$  0.284 & 170.94 &  1.62 &  1340 &   $\ldots$        &  $\ldots$ &  - &   3.14 $\pm$ 0.59 & $-$0.38 & 3 &   3.24 $\pm$ 0.36 & 13.84 & 4 &  3.64 $\pm$  0.28 & 5 &  3.14  $\pm$ 0.16 & 7 & 3.256 $\pm$ 0.135 & 3,5,7 \\
$\delta$ Cep &  $-1.172 \pm$  0.468 & 182.21 &  2.44 &  2100 &   $\ldots$        &  $\ldots$ &  - &   3.32 $\pm$ 0.58 &    0.41 & 3 &   3.77 $\pm$ 0.16 & -2.45 & 4 &  3.81 $\pm$  0.20 & 5 &  3.66  $\pm$ 0.15 & 7 & 3.723 $\pm$ 0.095 & 3,4,5,7 \\
FF Aql       &   1.810 $\pm$  0.107 &  65.83 &  0.49 &   100 &   1.64 $\pm$ 0.89 &  3.14     &  2 &   1.32 $\pm$ 0.72 &    0.43 & 3 &   2.11 $\pm$ 0.33 &  0.77 & 4 &  2.05 $\pm$  0.34 & 5 &  2.81  $\pm$ 0.18 & 7 & 2.543 $\pm$ 0.143 & 4,5,7 \\
$l$ Car      &   0.777 $\pm$  0.257 & 171.10 &  1.67 &  1190 &   $\ldots$        &  $\ldots$ &  - &   2.16 $\pm$ 0.47 & $-$0.49 & 3 &   2.09 $\pm$ 0.29 &  5.81 & 4 &  2.06 $\pm$  0.27 & 5 &  2.01  $\pm$ 0.20 & 7 & 2.042 $\pm$ 0.152 & 3,5,7 \\
RS Pup       &   0.584 $\pm$  0.026 &   7.74 &  0.00 &     0 &   0.63 $\pm$ 0.26 &  0.65     &  2 &   0.49 $\pm$ 0.68 & $-$0.73 & 3 &   1.91 $\pm$ 0.65 &  0.73 & 4 &  1.44 $\pm$  0.51 & 5 &  0.524 $\pm$ 0.022 & 6  & $\ldots$ & - \\ 
RT Aur       &   1.419 $\pm$  0.203 &  52.32 &  0.77 &   132 &   $\ldots$        & $\ldots$  &  - &   2.09 $\pm$ 0.89 & $-$0.05 & 3 & $-$1.10 $\pm$ 1.41 & 10.29 & 4 & -0.23 $\pm$  1.01 & 5 &  2.40  $\pm$ 0.19 & 7 &  2.40  $\pm$ 0.19 & 7 \\
SS CMa       &   0.201 $\pm$  0.029 &   4.37 &  0.00 &     0 &   0.69 $\pm$ 0.23 &  0.35     &  2 & $-$0.37 $\pm$ 1.75 &   1.32 & 3 &   0.40 $\pm$ 1.78 &  1.81 & 4 &  0.35 $\pm$  1.86 & 5 &  0.389 $\pm$ 0.029 & 9 & $\ldots$ & - \\  
 S Vul       &   0.305 $\pm$  0.041 &   7.98 &  0.00 &     0 & $-$0.21 $\pm$ 0.43 &  0.53    &  2 &   $\ldots$         & $\ldots$ & - & $\ldots$   & $\ldots$  & - & $\ldots$ & - &  0.322 $\pm$ 0.040 & 9  & $\ldots$ & - \\    
SY Aur       &   0.313 $\pm$  0.052 &   3.35 &  0.00 &     0 &   0.69 $\pm$ 0.25 &  0.48     &  2 &   1.15 $\pm$  1.70 &   0.27 & 3 & $-$1.84 $\pm$ 1.72 &  1.31 & 4 & -0.52 $\pm$  1.44 & 5 &  0.428 $\pm$ 0.054 & 8  & $\ldots$ & - \\ 
 T Vul       &   1.674 $\pm$  0.089 &  44.55 &  0.33 &    56 &   $\ldots$        & $\ldots$  &  - &   1.95 $\pm$  0.60 & $-$0.24 & 3 &   2.71 $\pm$ 0.43 &  1.37 & 4 &  2.31 $\pm$  0.29 & 5 &  1.90  $\pm$ 0.23 & 7 &  2.156 $\pm$ 0.166&  4,5,7 \\
VX Per       &   0.330 $\pm$  0.031 &   3.81 &  0.00 &     0 &   0.50 $\pm$ 0.27 &  0.71     &  2 &   1.08 $\pm$  1.48 &   0.04  & 3 &   0.87 $\pm$ 1.52 &  1.07 & 4 &  1.10 $\pm$  1.62 & 5 &  0.420 $\pm$ 0.074 & 9  & $\ldots$ & - \\  
VY Car       &   0.512 $\pm$  0.041 &   1.64 &  0.00 &     0 &   0.73 $\pm$ 0.29 &  0.62     &  2 &   1.28 $\pm$  1.76 &   2.88  & 3 &   0.36 $\pm$ 1.42 &  4.91 & 4 &  1.56 $\pm$  0.91 & 5 &  0.586 $\pm$ 0.044 & 9  & $\ldots$ & - \\  
 W Sgr       &   1.180 $\pm$  0.412 &  88.20 &  1.40 &   371 &   $\ldots$        & $\ldots$  &  - &   1.57 $\pm$  0.93 &   0.47  & 3 &   3.75 $\pm$ 1.12 & 10.40 & 4 &  2.59 $\pm$  0.75 & 5 &  2.28  $\pm$ 0.20 & 7 & 2.28  $\pm$ 0.20 &  7 \\
WZ Sgr       &   0.513 $\pm$  0.077 &   3.52 &  0.00 &     0 &   $\ldots$        & $\ldots$  &  - & $-$0.75 $\pm$ 1.76 & $-$0.40 & 3 &   3.50 $\pm$ 1.22 & $-$0.12 & 4 &  2.46 $\pm$  1.12 & 5 &  0.512 $\pm$ 0.037 & 9 & $\ldots$ & - \\ 
 X Pup       &   0.302 $\pm$  0.043 &   1.20 &  0.00 &     0 &   0.28 $\pm$ 0.29 &  0.58     &  2 & $-$0.05 $\pm$ 1.10 &   1.31  & 3 &   1.97 $\pm$ 1.26 & $-$0.82 & 4 &  2.87 $\pm$  0.92 & 5 &  0.277 $\pm$ 0.047 & 9  & $\ldots$ & -  \\ 
 X Sgr       &   3.431 $\pm$  0.202 &  73.61 &  0.78 &   151 &   $\ldots$        & $\ldots$  &  - &   3.03 $\pm$  0.94 &   0.63  & 3 &   3.31 $\pm$ 0.26 & $-$0.63 & 4 &  3.39 $\pm$  0.21 & 5 &  3.00  $\pm$ 0.18 & 7  & 3.197 $\pm$ 0.121 & 4,5,7 \\
XY Car       &   0.330 $\pm$  0.027 &   7.50 &  0.00 &     0 &   0.19 $\pm$ 0.24 &  0.50     &  2 & $-$0.62 $\pm$ 0.95 & $-$0.05 & 3 & $-$1.02 $\pm$ 0.88 &  0.18 & 4 & $-$0.75 $\pm$  0.87 & 5 &  0.438 $\pm$ 0.047 & 9  & $\ldots$ & - \\  
 Y Sgr       &  $-0.470 \pm$  0.280 &  73.09 &  0.75 &   143 &   $\ldots$        &  $\ldots$ &  - &   2.52 $\pm$  0.93 & $-$2.19 & 3 &   2.64 $\pm$ 0.45 & $-$0.92 & 4 &  3.73 $\pm$  0.32 & 5 &  2.13  $\pm$ 0.29 & 7 & 2.812 $\pm$ 0.194 & 4,5,7 \\
$\zeta$ Gem  &   2.250 $\pm$  0.301 &  90.10 &  1.16 &   389 &   $\ldots$        &  $\ldots$ &  - &   2.79 $\pm$  0.81 & $-$0.18 & 3 &   2.37 $\pm$ 0.30 &    1.19 & 4 &  2.71 $\pm$  0.17 & 5 &  2.78  $\pm$ 0.18 & 7  & 2.689 $\pm$ 0.114 & 4,5,7 \\

\hline
\end{tabular}

\tablefoot{
Columns 2-5: Parallax (error), GOF, excess noise parameter, and significance from GDR2.
Columns 6 and 7: Parallax (error) and excess noise from GDR1 (Reference 2). 
Columns 9 and 10: Parallax (error) and GOF parameter from {\it Hipparcos} (Reference 3, \citealt{ESA1997}).
Columns 12 and 13: Parallax (error) and GOF parameter from {\it Hipparcos} (Reference 4, \citealt{vanL07NR}).
Column 15: Parallax (error) from {\it Hipparcos} (Reference 5, \citealt{vanL07}).
Column 17: Parallax (error) from other sources.
References:  6= The parallax for RS Pup is from \citet{Kervella2014} and is a geometric distance based on the light echo seen in the 
nebula surrounding the star,
7= based on {\it HST-FGS}   parallaxes from \citet{Benedict07},
8= based on {\it HST-WFC3}  parallax   from \citet{Riess14}, 
9= based on {\it HST-WFC3}  parallaxes from \citet{Riess18},
10= based on {\it HST-FGS}  parallax for Polaris~B from \citet{Bond18}.
Column 19: weighted mean parallax (error) from the sources listed in Column 20.
}

\label{Tab:Hipp}
\end{sidewaystable*}

\section{Analysis} 
\label{S-Ana}

The fundamental equation between parallax, apparent, and absolute magnitude is
\begin{equation}
  \pi = 100 \cdot 10^{0.2 \; (M-m)},
\end{equation}
where $\pi$ is the parallax in mas, and $m$ the dereddened apparent magnitude. 
The absolute magnitude $M$ is parameterised as
\begin{equation}
  M = \alpha + \beta \, \log P + \gamma \,  {\rm [Fe/H]}
\end{equation}
and the aim is to derive the coefficients $\alpha$, $\beta$, and $\gamma$.

\citet{FC1997} had a similar aim in mind using \Hp\ data. The accuracy of the \Hp\ data was such that only 
the simpler problem with $\gamma = 0$ and known slope $\beta$ (from Cepheids in the Large Magellanic Cloud, LMC) could be tackled.
In that case the problem is simplified to $10^{0.2 \alpha} = 0.01 \cdot \pi \cdot 10^{0.2 \; (m - \beta \, \log P)}$.
The zero point of the $PL$ relation was found by taking the weighted mean of the term on the right-hand side over all 223 Cepheids 
available to them, and then calculating 5 times the logarithmic value.
For an assumed slope of $-2.81$ in the $V$ band they derived $\alpha = -1.43 \pm 0.10$ mag.
Using the revised \Hp\ parallaxes \citet{vanL07} found  $\alpha = -2.47 \pm 0.03$ mag (the weighted mean of the three values in their Table~6) 
for fixed $\beta = -3.26$ in the $K$ band, see Table~\ref{Tab:plfits}.

In  the present paper, in principle, we want to solve for all coefficients and therefore the non-linear problem of fitting Eq.~1 
is solved directly using the Levenberg--Marquardt algorithm (as implemented in Fortran in Numerical Recipes, \citealt{Press1992}).

An important advantage of using Eq.~1 in this form is that no selection on positive parallaxes or relative parallax error is required, and therefore 
the results are not subject to Lutz--Kelker bias (Lutz \& Kelker 1973) (see the discussion in \citealt{FC1997, Koen1998, Lanoix1999}).
It is also one of the methods, known as  astrometry-based luminosity (ABL), used in \citet{Clementini17} to analyse GDR1 data (also see \citealt{Luri18}) .
Another advantage is that the errors in the parallax can be assumed to be symmetric and Gaussian distributed.

Monte Carlo simulations are carried out for an improved understanding of the results.
The basic data of the Cepheids (Table~\ref{Tab:Targets}) are read in, together with the parameters of the \G\ DR2 (or the external parallax data).
Periods of FO pulsators (type DCEPS) are fundamentalised using  $P_0 = P_1/(0.716-0.027\, \log P_1)$ following \citet{FC1997}.
Then,

\begin{itemize}
\item A new parallax is drawn from a Gaussian with the adopted mean and error. A parallax zero-point offset may be applied;

\item A new period is drawn from a Gaussian with the mean  input period and an error equal to $1.3 \cdot 10^{-4} P$;

\item A new [Fe/H] is drawn from a Gaussian with the mean  input iron abundance and an error of 0.08 dex;
 
\item A new reddening is drawn from a Gaussian with the mean and error from the input. Negative reddenings are set to zero;

\item The input $V, J, H, K$ magnitudes are dereddened (see Sect.~\ref{S-PRE} for details);

\item New $V, J, H, K$ magnitudes are drawn from Gaussians using the dereddened magnitudes and assumed error bars 
(see Sect.~\ref{S-PRE} for details);

\item The NIR magnitudes are transformed to the 2MASS system if needed;

\item To take into account the intrinsic width in the instability strip (see \citealt{FC1997}) the value of $(M-m)$ is
increased by a value drawn from a Gaussian centred on 0 with error $\delta_{\rm PL}$. 

\end{itemize}

\noindent
The  generated data ($\pi$, $\spi$, magnitude, period, colour, iron abundance) are fed to the  minimization routine that returns
the parameters ($\alpha$, $\beta$, $\gamma$) with internal error bars.
This is repeated $N$ times (typically $N= $1001). 
The best estimate of the parameters is taken as the median of its distribution with the $1 \sigma$ error bar derived from 
the 2.7\% and 93.7\% percentiles ($\pm 2 \sigma$). This results in realistic error bars that takes into account all likely variations in the data.

Using the best fit parameters ($\alpha, \beta, \gamma$ for each simulation) the photometric parallax is calculated  using Eqs.~1 and 2 
and compared to the observed parallax. This gives a standard deviation per star and when these values are ordered over all simulations, stars 
that are systematic outliers can be identified. An rms deviation of the fit to the data is also calculated.

\section{Simulations}
\label{S-Sim}

To test the methodology and the sensitivity to the parameters, simulations were carried out.
Parallax and error in the parallax are generated based on the period, reddening, magnitudes of the stars in the sample, and
a period-luminosity relation.
These observed parameters are then fed to the code and analysed as described above.
The results are listed in Table~\ref{Tab:plfitsSim}.
The sample size was chosen to represent the number of FU and FO pulsators in the sample.
Solutions 1--4 are idealised: the simulated parallax is exactly 1/distance, and the error in the parallax is based on a fraction of the distance.
The input slope and zero point are retrieved almost exactly.
In all other simulations the simulated parallax is based on the exact parallax plus a Gaussian distributed error.
In solutions 5--8 this is still based on a hypothetical fractional error in the distance.
In all other simulations the error in the parallax is based on the observed distribution among the stars with a GOF $<8$.
The distribution is not a strong function of $G$-band (or $V$-band) magnitude in the present sample, and $\spi$ is represented 
by a mean of 0.040 and a Gaussian dispersion of 0.013 with a minimum value of 0.020 mas.
Solution 9 is the standard case which returns the input slope and zero point with a slight offset, but still within the error bars. 
Selecting on parallax error does not influence this (solutions~12 and 13. We note that in the present sample,  very 
few or no negative parallaxes are predicted).

The last part of the simulations investigates the influence of a zero-point offset in the parallax. 
The all-sky average value derived from QSOs ($-0.029$ mas, \citealt{Lindegren18}) and for a sample of 50 CCs ($-0.046$ mas, \citealt{RiessGDR2}) are used to illustrate this.

The impact is significant on both zero point and slope of the $PL$ relation when not considered in the analysis of the data (solutions~14 and 15). 
The true solution can be partially recovered when the parallax zero-point offset is partially taken into account in the analysis (solutions~16 and 17). 
Making a strong selection on the relative parallax error or parallax may also recover an unrecognised parallax zero-point offset, but 
at the expense of a larger error bar in the zero point and slope of the $PL$ relation (solutions~18--21 and 22--24).

The last two entries are for a realistic sample size (see next section),  also including an intrinsic width of the instability strip (IS).
In both cases this leads to larger error bars in the derived parameters.

\begin{table*} 

\caption{Simulations of the $PL$ relations of the form $M = \alpha + \beta \log P$ with input values $\alpha= -2.50$ and $\beta = -3.30$; 
$N$ is the number of stars in the solution.}

\begin{tabular}{rccrl} \hline \hline 

\#  &      $\alpha$    &       $\beta$     & $N$     & Remarks \\ 
\hline 

\multicolumn{4}{c}{Using exact distance}  \\

1 & -2.502 $\pm$ 0.006 & -3.296 $\pm$ 0.007 & 426 &  1\% error on distance      \\
2 & -2.502 $\pm$ 0.009 & -3.296 $\pm$ 0.011 & 426 &  2\% error on distance      \\
3 & -2.502 $\pm$ 0.018 & -3.296 $\pm$ 0.020 & 426 &  5\% error on distance      \\
4 & -2.504 $\pm$ 0.037 & -3.294 $\pm$ 0.041 & 426 & 10\% error on distance      \\

\multicolumn{4}{c}{With parallax error}  \\
5 & -2.499 $\pm$ 0.006 & -3.297 $\pm$ 0.007 & 426 &  1\% error on distance      \\
6 & -2.500 $\pm$ 0.009 & -3.294 $\pm$ 0.011 & 426 &  2\% error on distance      \\
7 & -2.496 $\pm$ 0.018 & -3.293 $\pm$ 0.020 & 426 &  5\% error on distance      \\
8 & -2.488 $\pm$ 0.037 & -3.294 $\pm$ 0.041 & 426 & 10\% error on distance      \\ 

9  & -2.521 $\pm$ 0.021 & -3.267 $\pm$ 0.022 & 426 & parallax error based on data      \\ 

10 & -2.518 $\pm$ 0.027 & -3.268 $\pm$ 0.030 & 426 & error based on data, another random seed      \\ 
11 & -2.514 $\pm$ 0.023 & -3.273 $\pm$ 0.025 & 426 & error based on data, another random seed  \\ 
12 & -2.511 $\pm$ 0.021 & -3.276 $\pm$ 0.023 & 361 & error based on data, $\pi >0, \spi/\pi<0.2$     \\ 
13 & -2.514 $\pm$ 0.024 & -3.273 $\pm$ 0.027 & 218 & error based on data, $\pi >0, \spi/\pi<0.1$     \\ 

14 & -2.557 $\pm$ 0.022 & -3.295 $\pm$ 0.024 & 426 & error based on data, ZPoff= $-0.029$ mas, not in analysis    \\ 
15 & -2.577 $\pm$ 0.022 & -3.312 $\pm$ 0.024 & 426 & error based on data, ZPoff= $-0.046$ mas, not in analysis         \\ 
16 & -2.521 $\pm$ 0.022 & -3.267 $\pm$ 0.023 & 426 & error based on data, ZPoff= $-0.046$ mas, $-0.046$ in analysis    \\ 
17 & -2.541 $\pm$ 0.022 & -3.284 $\pm$ 0.023 & 426 & error based on data, ZPoff= $-0.046$ mas, $-0.029$ in analysis    \\ 

18 & -2.565 $\pm$ 0.023 & -3.319 $\pm$ 0.025 & 310 & error based on data, ZPoff= $-0.046$ mas, not in analysis, $\pi >0, \spi/\pi<0.2$   \\ 
19 & -2.562 $\pm$ 0.023 & -3.308 $\pm$ 0.024 & 189 & error based on data, ZPoff= $-0.046$ mas, not in analysis, $\pi >0, \spi/\pi<0.1$   \\ 
20 & -2.572 $\pm$ 0.025 & -3.271 $\pm$ 0.027 &  76 & error based on data, ZPoff= $-0.046$ mas, not in analysis, $\pi >0, \spi/\pi<0.05$   \\ 
21 & -2.542 $\pm$ 0.040 & -3.276 $\pm$ 0.038 &  13 & error based on data, ZPoff= $-0.046$ mas, not in analysis, $\pi >0, \spi/\pi<0.02$   \\ 

22 & -2.575 $\pm$ 0.026 & -3.273 $\pm$ 0.028 & 126 & error based on data, ZPoff= $-0.046$ mas, not in analysis, $\pi >0.5$ mas  \\ 
23 & -2.570 $\pm$ 0.030 & -3.257 $\pm$ 0.033 &  50 & error based on data, ZPoff= $-0.046$ mas, not in analysis, $\pi >1.0$ mas   \\ 
24 & -2.539 $\pm$ 0.035 & -3.279 $\pm$ 0.035 &  20 & error based on data, ZPoff= $-0.046$ mas, not in analysis, $\pi >1.5$ mas   \\ 

\\

25 & -2.515 $\pm$ 0.033 & -3.288 $\pm$ 0.039 & 205 & as (9), for realistic sample size and $\delta_{\rm PL}= 0$ mag     \\ 
26 & -2.520 $\pm$ 0.058 & -3.289 $\pm$ 0.066 & 205 & as (9), for realistic sample size and $\delta_{\rm PL}= 0.066$ mag \\ 

\hline 

\end{tabular} 
\label{Tab:plfitsSim}
\end{table*}

\section{Results}
\label{S-Res}
\subsection{$PL$ relation}

Two sources of parallax data are considered. The first is exclusively based on GDR2 data. 
However, many bright stars have a very poor astrometric solution and have to be discarded (see Table~\ref{Tab:Hipp} and the discussion below).
In the second sample the poorest GDR2 parallax data (with GOF $>8$) is replaced by external data when available.
These are the 11 stars with an entry in the last column of Table~\ref{Tab:Hipp} (as having a non-\Hp\ based accurate parallax) and 
9 stars where the weighted mean of the available \Hp\ parallaxes was used\footnote{Specifically: AH Vel 1.454 $\pm$ 0.174 mas, 
BG Cru 2.282 $\pm$ 0.209 mas, DT Cyg 2.060 $\pm$ 0.233 mas, $\eta$ Aql 3.150 $\pm$ 0.579 mas, MY Pup 1.097 $\pm$ 0.143 mas, 
SU Cas 2.549 $\pm$ 0.229 mas,  R Cru 2.060 $\pm$ 0.480 mas, and  S Mus 2.120 $\pm$ 0.350 mas. 
Only \Hp\ parallax solutions with a GOF $<3.5$ and relative parallax error below 25\% were considered in the averaging.}.

For the moment we do not consider the metallicity and concentrate on the classical $PL$ relation in the $V$, $K$, and $WVK$ bands, 
the Wesenheit reddening-free index defined as $WVK= K - 0.13\;(V-K)$ \citep{Ripepi12}, based on the reddening law of \citet{Cardelli89}.
A large number of models was run to investigate the influence of FU versus FO pulsators, period cuts, selection on GOF, 
the parallax zero-point offset, the intrinsic width of the IS, and to identify systematic outliers.
For completeness the model results are reported and briefly discussed in Appendix~\ref{AppIni}. 
The conclusions are to conservatively use only FU mode pulsators (the FO mode pulsators seem to show a significantly different slope), 
consider stars between 2.7 and 35 days (mainly for the comparison to SMC and LMC results), 
and select stars with $\mid$GOF$\mid<8$ and $\epsilon_i<0.001$. 
The stars U Vul, V Vel, V636 Cas, V526 Aql, QQ Per, and HQ Car are also excluded. The last two are likely T2C.
The standard values for $\delta_{\rm PL}$ of 0.20, 0.066, and 0.049 mag in the $V$, $K,$ and $WVK$ bands are adopted below \citep{Inno16}.

Table~\ref{Tab:plfits} lists the results for this reduced sample of stars. Solutions 1--9 using only GDR2 parallaxes (marked `GDR2' in the table), 
solutions 10--12 using external parallaxes when available for the poorest GDR2 data (marked `GDR2+Ext'. Any parallax zero-point offset is only 
applied to the GDR2 objects). 
In these first solutions zero point and slope of the $PL$ relations are both fitted. For a fixed parallax zero-point offset adding the 
external parallax data does not have a very big effect (less then 1$\sigma$; solutions 1, 4, 7 versus 10, 11, 12). 

The effect of a parallax zero-point offset is larger, and systematic.
Increasing the \G\ parallax (making the offset more negative in the convention used by \citealt{Lindegren18}) makes the slope less steep and 
the zero point fainter in all three filters considered.

The slope of the $PL$ relation derived in the $V$, $K$, and $WVK$ filters is different than that in the LMC.
Table~\ref{Tab:plfitsLit} compiles recent determinations of the $PL(Z)$ relation in these filters in the SMC, LMC, and the Milky Way.
They include the zero point and slope derived for Galactic Cepheids using \Hp\ data \citep{FC1997,vanL07}, the results by \citet{Fouque07} that 
have been used in the analysis of GDR1 data in \citet{Clementini17}.
Listed next are the results by \citet{Storm11a,Storm11b} and \citet{Gr2013} from Baade--Wesselink distances to SMC, LMC, and Galactic objects.
The solutions for the Galactic and LMC objects are listed, and the solutions for all objects, which then includes a few CCs in the SMC as well.
The last sets of entries are based on OGLE data in the $V$ band, sometimes combined with sets of NIR photometry.
We note  that there is no single slope (and zero point) for the LMC (and SMC) Cepheids. The values depend on the period cut used, but also
on more seemingly subtle issues. \citet{Inno16} discuss the differences with \citet{Jacyszyn16} which both use the OGLE-IV catalogue of CCs.
In the $I$ band the papers find different slopes ($-3.327 \pm 0.001$ versus $-3.313 \pm 0.006$) and \citet{Inno16} trace this back to a
difference in the sigma-clipping procedure (removing 3$\sigma$ versus 6$\sigma$ outliers).

In the $V$ band the difference in slope between the LMC ($\sim -2.65$ to $-2.8$) and the GDR2 results ($\sim -2.35$ to $-2.38$) is at the 3--4$\sigma$ level, 
and larger when applying a parallax zero-point offset.
On the other hand, of the three bands considered the $V$ band is the one where the evidence for non-linearity is strongest and where the slope
of SMC Cepheids is known to be inconsistent with that of LMC Cepheids \citep{Ngeow09, Ngeow15}.
In the $K$ band the difference in slope between the LMC ($\sim -3.20$ to $-3.35$) and the GDR2 results ($\sim -3.10$ to $-3.15$) is at the 1--2$\sigma$ level, 
but becomes significant when applying a parallax zero-point offset.
In the $WVK$ band the same tendency is seen, the difference in slope between the LMC and the GDR2 results becomes significant when applying a parallax zero-point offset.

In solution 20 and  in Table~\ref{Tab:plfits} the slope of the $PL$ relation has been fixed to different values found for LMC Cepheids.
This has been done for both sources of parallax data, and for different parallax zero-point offsets.
For a fixed slope, the difference between the zero points gives the distance modulus (DM) to the LMC, and this is listed in the last column for selected models.
The quoted error considers the error in both the Galactic and LMC zero points.
The DM depend strongly on the adopted parallax zero-point offset. For an offset of $-0.046$ mas the DM is essentially 18.70 mag based on the $K$ and $WVK$ band
(it is 0.1 mag shorter but with a larger error bar in $V$). Adding the external parallax data makes these distances shorter by about 0.02 mag.
The current analysis suggests that a DM of $18.493 \pm 0.048$ mag (based on the eclipsing binaries; \citealt{Pietrzynski13}) or the recommended value of
18.49 $\pm$ 0.09 mag (based on several independent distance indicators; \citealt{deGB2014}) requires a significantly larger offset (in absolute sense) 
of order  $-0.1$ mas. This is comparable to the $-0.118 \pm 0.003$ listed in \citet{Arenou18} for the difference between GDR2 and \Hp\ parallaxes, 
but larger (in absolute sense) than given by any of the other comparisons listed in their Table~1.

\begin{table*} 

\caption{$PL$ relations of the form $M = \alpha + \beta \log P$;
$N$ is the number of stars in the solution.}

\begin{tabular}{rrrrll} \hline \hline 

 &  $\alpha$   & $\beta$    & $N$ &  Constraints  & LMC DM  \\  
 &             &            &     &                & (mag) \\
\hline 

\multicolumn{5}{r}{ FU, $2.7 < P(d) < 35$, $\mid$GOF$\mid<8$, $\epsilon_i<0.001$, $\delta_{\rm PL}$ applied. 
} \\

1 & -1.919 $\pm$ 0.119 & -2.386 $\pm$ 0.138 & 194 & V, GDR2, ZPoff= $0$ mas                 \\ 
2 & -1.875 $\pm$ 0.118 & -2.305 $\pm$ 0.136 & 194 & V, GDR2, ZPoff= $-0.029$ mas  \\ 
3 & -1.848 $\pm$ 0.119 & -2.260 $\pm$ 0.135 & 194 & V, GDR2, ZPoff= $-0.046$ mas  \\ 

4 & -2.912 $\pm$ 0.058 & -3.154 $\pm$ 0.070 & 194 & K, GDR2, ZPoff= $0$ mas                   \\ 
5 & -2.866 $\pm$ 0.057 & -3.071 $\pm$ 0.068 & 194 & K, GDR2, ZPoff= $-0.029$ mas   \\ 
6 & -2.839 $\pm$ 0.056 & -3.028 $\pm$ 0.067 & 194 & K, GDR2, ZPoff= $-0.046$ mas   \\ 

7 & -3.047 $\pm$ 0.055 & -3.252 $\pm$ 0.066 & 194 & WVK, GDR2, ZPoff= $0$ mas                   \\ 
8 & -2.999 $\pm$ 0.053 & -3.170 $\pm$ 0.063 & 194 & WVK, GDR2, ZPoff= $-0.029$ mas   \\ 
9 & -2.972 $\pm$ 0.052 & -3.126 $\pm$ 0.063 & 194 & WVK, GDR2, ZPoff= $-0.046$ mas   \\ 

\\

10 & -1.917 $\pm$ 0.118 & -2.351 $\pm$ 0.137 & 205 & V, GDR2+Ext, ZPoff= $0$ mas    \\ 
11 & -2.908 $\pm$ 0.057 & -3.109 $\pm$ 0.068 & 205 & K, GDR2+Ext, ZPoff= $0$ mas    \\ 
12 & -3.041 $\pm$ 0.053 & -3.207 $\pm$ 0.063 & 205 & WVK, GDR2+Ext, ZPoff= $0$ mas  \\ 

\\

20 & -1.728 $\pm$ 0.029 & -2.629 fixed & 194 & V, GDR2, ZPoff= $0$ mas                  \\ 
21 & -1.619 $\pm$ 0.029 & -2.629 fixed & 194 & V, GDR2, ZPoff= $-0.029$ mas  \\ 
22 & -1.557 $\pm$ 0.029 & -2.629 fixed & 194 & V, GDR2, ZPoff= $-0.046$ mas  \\ 

23 & -1.690 $\pm$ 0.029 & -2.678 fixed & 194 & V, GDR2, ZPoff= $0$ mas                  \\ 
24 & -1.581 $\pm$ 0.030 & -2.678 fixed & 194 & V, GDR2, ZPoff= $-0.029$ mas  \\ 
25 & -1.519 $\pm$ 0.030 & -2.678 fixed & 194 & V, GDR2, ZPoff= $-0.046$ mas  \\ 

26 & -1.589 $\pm$ 0.030 & -2.810 fixed & 194 & V, GDR2, ZPoff= $0$ mas                & 18.761 $\pm$ 0.030 \\ 
27 & -1.480 $\pm$ 0.030 & -2.810 fixed & 194 & V, GDR2, ZPoff= $-0.029$ mas & 18.650 \\ 
28 & -1.418 $\pm$ 0.030 & -2.810 fixed & 194 & V, GDR2, ZPoff= $-0.046$ mas & 18.590 \\ 
29 & -1.321 $\pm$ 0.030 & -2.810 fixed & 194 & V, GDR2, ZPoff= $-0.074$ mas & 18.493 \\ 
30 & -1.233 $\pm$ 0.030 & -2.810 fixed & 194 & V, GDR2, ZPoff= $-0.100$ mas & 18.405 \\ 
 
40 & -2.879 $\pm$ 0.014 & -3.194 fixed & 194 & K, GDR2, ZPoff= $0$ mas                 & 18.875 $\pm$ 0.017 \\ 
41 & -2.769 $\pm$ 0.014 & -3.194 fixed & 194 & K, GDR2, ZPoff= $-0.029$ mas & 18.765  \\ 
42 & -2.707 $\pm$ 0.014 & -3.194 fixed & 194 & K, GDR2, ZPoff= $-0.046$ mas & 18.703  \\ 

43 & -2.827 $\pm$ 0.014 & -3.260 fixed & 194 & K, GDR2, ZPoff= $0$ mas                 & 18.880 $\pm$ 0.014 \\ 
44 & -2.717 $\pm$ 0.014 & -3.260 fixed & 194 & K, GDR2, ZPoff= $-0.029$ mas & 18.770  \\ 
45 & -2.655 $\pm$ 0.014 & -3.260 fixed & 194 & K, GDR2, ZPoff= $-0.046$ mas & 18.708  \\ 
46 & -2.469 $\pm$ 0.013 & -3.260 fixed & 194 & K, GDR2, ZPoff= $-0.100$ mas & 18.522  \\ 

47 & -2.800 $\pm$ 0.014 & -3.295 fixed & 194 & K, GDR2, ZPoff= $0$ mas                       & 18.870 $\pm$ 0.022 \\ 
48 & -2.690 $\pm$ 0.014 & -3.295 fixed & 194 & K, GDR2, ZPoff= $-0.029$ mas & 18.760  \\ 
49 & -2.628 $\pm$ 0.014 & -3.295 fixed & 194 & K, GDR2, ZPoff= $-0.046$ mas  & 18.698  \\ 
50 & -2.442 $\pm$ 0.013 & -3.295 fixed & 194 & K, GDR2, ZPoff= $-0.100$ mas  & 18.512  \\ 
51 & -2.377 $\pm$ 0.013 & -3.295 fixed & 194 & K, GDR2, ZPoff= $-0.120$ mas  & 18.447  \\ 

52 & -2.745 $\pm$ 0.014 & -3.365 fixed & 194 & K, GDR2, ZPoff= $0$ mas                   \\ 
52 & -2.636 $\pm$ 0.014 & -3.365 fixed & 194 & K, GDR2, ZPoff= $-0.029$ mas    \\ 
53 & -2.574 $\pm$ 0.014 & -3.365 fixed & 194 & K, GDR2, ZPoff= $-0.046$ mas    \\ 

60 & -2.997 $\pm$ 0.013 & -3.314 fixed & 194 & WVK, GDR2, ZPoff= $0$ mas                 \\ 
61 & -2.887 $\pm$ 0.013 & -3.314 fixed & 194 & WVK, GDR2, ZPoff= $-0.029$ mas  \\ 
62 & -2.825 $\pm$ 0.013 & -3.314 fixed & 194 & WVK, GDR2, ZPoff= $-0.046$ mas  \\ 
63 & -2.793 $\pm$ 0.013 & -3.314 fixed & 194 & WVK, GDR2, ZPoff= $-0.055$ mas  \\ 

64 & -2.988 $\pm$ 0.013 & -3.325 fixed & 194 & WVK, GDR2, ZPoff= $0$ mas              & 18.858  $\pm$ 0.018 \\ 
65 & -2.878 $\pm$ 0.013 & -3.325 fixed & 194 & WVK, GDR2, ZPoff= $-0.029$ mas  & 18.748 \\ 
66 & -2.816 $\pm$ 0.013 & -3.325 fixed & 194 & WVK, GDR2, ZPoff= $-0.046$ mas  & 18.696 \\ 
67 & -2.784 $\pm$ 0.012 & -3.325 fixed & 194 & WVK, GDR2, ZPoff= $-0.055$ mas  & 18.654 \\ 
68 & -2.714 $\pm$ 0.012 & -3.325 fixed & 194 & WVK, GDR2, ZPoff= $-0.075$ mas  & 18.584 \\ 
69 & -2.630 $\pm$ 0.012 & -3.325 fixed & 194 & WVK, GDR2, ZPoff= $-0.100$ mas  & 18.500 \\ 

\\

70 & -1.544 $\pm$ 0.029 & -2.629 fixed & 205 & V, GDR2+Ext, ZPoff= $-0.046$ mas   \\ 
71 & -1.506 $\pm$ 0.030 & -2.678 fixed & 205 & V, GDR2+Ext, ZPoff= $-0.046$ mas   \\ 
72 & -1.404 $\pm$ 0.030 & -2.810 fixed & 205 & V, GDR2+Ext, ZPoff= $-0.046$ mas  & 18.576 $\pm$ 0.030 \\ 

73 & -2.684 $\pm$ 0.013 & -3.194 fixed & 205 & K, GDR2+Ext, ZPoff= $-0.046$ mas   \\ 
74 & -2.632 $\pm$ 0.013 & -3.260 fixed & 205 & K, GDR2+Ext, ZPoff= $-0.046$ mas  & 18.685  $\pm$ 0.013 \\ 
75 & -2.604 $\pm$ 0.013 & -3.295 fixed & 205 & K, GDR2+Ext, ZPoff= $-0.046$ mas   \\ 
76 & -2.550 $\pm$ 0.013 & -3.365 fixed & 205 & K, GDR2+Ext, ZPoff= $-0.046$ mas   \\ 

77 & -2.800 $\pm$ 0.012 & -3.314 fixed & 205 & WVK, GDR2+Ext, ZPoff= $-0.046$ mas   \\ 
78 & -2.792 $\pm$ 0.012 & -3.325 fixed & 205 & WVK, GDR2+Ext, ZPoff= $-0.046$ mas  & 18.662 $\pm$ 0.018 \\ 

\hline

\end{tabular} 
\label{Tab:plfits}
\end{table*}

\begin{table*} 

\caption{$PL(Z)$ relations in the literature of the form $M = \alpha + \beta \log P + \gamma$ [Fe/H]
for different galaxies. }

\begin{tabular}{lcrrrrl} \hline \hline 

Band    & Galaxy & N & $\alpha$ & $\beta$ & $\gamma$ &  Remarks \\  
\hline

 \multicolumn{6}{c}{ \citet{FC1997} } \\
  V & GAL &  220  & -1.43 $\pm$ 0.10 & -2.81 fixed  & & Hipparcos \\

 \multicolumn{6}{c}{ \citet{vanL07} } \\
  K & GAL &  220  & -2.47 $\pm$ 0.03 & -3.26 fixed  & & Hipparcos \\

 \multicolumn{6}{c}{ \citet{Fouque07} } \\
  V & GAL & 58  & -1.275 $\pm$ 0.023 & -2.678 $\pm$ 0.076 &  \\ 
  K & GAL & 58  & -2.282 $\pm$ 0.019 & -3.365 $\pm$ 0.063 &  \\ 

 \multicolumn{6}{c}{ \citet{Clementini17}, ABL solutions } & GDR1 \\
  V & GAL & 102   & -1.54 $\pm$ 0.10 & -2.678 fixed &  \\
  K & GAL & 102   & -2.63 $\pm$ 0.10 & -3.365 fixed &  \\
WVK & GAL & 102   & -2.87 $\pm$ 0.10 & -3.32  fixed &   \\ 

\multicolumn{6}{c}{ \citet{Storm11a,Storm11b} } & Baade--Wesselink distances \\

  V & GAL &  70   & -1.29 $\pm$ 0.03 & -2.67 $\pm$ 0.10 & \\
  K & GAL &  70   & -2.33 $\pm$ 0.03 & -3.33 $\pm$ 0.09 & \\

  V & LMC &  36   & -1.22 $\pm$ 0.03 & -2.78 $\pm$ 0.11 & \\
  K & LMC &  36   & -2.36 $\pm$ 0.04 & -3.28 $\pm$ 0.09 & \\

  V & ALL & 111   & -1.24 $\pm$ 0.03 & -2.73 $\pm$ 0.07 & +0.09 $\pm$ 0.10 \\
  K & ALL & 111   & -2.35 $\pm$ 0.02 & -3.30 $\pm$ 0.06 & $-0.11$ $\pm$ 0.10 \\

\multicolumn{6}{c}{\citet{Gr2013}} & Baade--Wesselink distances  \\

V & ALL & 160 & -1.48 $\pm$ 0.08 & -2.40 $\pm$ 0.07 &  \\
V & GAL & 119 & -1.68 $\pm$ 0.10 & -2.21 $\pm$ 0.09 &  \\
V & LMC &  36 & -1.10 $\pm$ 0.17 & -2.69 $\pm$ 0.12 &  \\

V & ALL & 160 & -1.55 $\pm$ 0.09 & -2.33 $\pm$ 0.07 & +0.23 $\pm$ 0.11 \\ 
V & GAL & 121 & -1.69 $\pm$ 0.10 & -2.21 $\pm$ 0.09 & +0.17 $\pm$ 0.25 \\
V & LMC &  36 & -1.09 $\pm$ 0.17 & -2.68 $\pm$ 0.12 & $-0.14$ $\pm$ 0.35 \\

K & ALL & 162 & -2.50 $\pm$ 0.08 & -3.06 $\pm$ 0.06 &  \\
K & GAL & 121 & -2.55 $\pm$ 0.09 & -3.03 $\pm$ 0.08 &  \\
K & LMC &  36 & -2.26 $\pm$ 0.17 & -3.21 $\pm$ 0.13 &  \\

K & ALL & 162 & -2.49 $\pm$ 0.08 & -3.07 $\pm$ 0.07 & $-0.05$ $\pm$ 0.10 \\ 
K & GAL & 121 & -2.56 $\pm$ 0.09 & -3.03 $\pm$ 0.08 & +0.07 $\pm$ 0.20 \\
K & LMC &  36 & -2.27 $\pm$ 0.18 & -3.22 $\pm$ 0.13 & +0.19 $\pm$ 0.37 \\

WVK & ALL & 158 & -2.68 $\pm$ 0.08 & -3.12 $\pm$ 0.06 &  \\
WVK & GAL & 120 & -2.69 $\pm$ 0.09 & -3.12 $\pm$ 0.08 &  \\
WVK & LMC &  36 & -2.41 $\pm$ 0.18 & -3.27 $\pm$ 0.13 &  \\

WVK & ALL & 158 & -2.69 $\pm$ 0.08 & -3.11 $\pm$ 0.07 & +0.04 $\pm$ 0.10 \\ 
WVK & GAL & 120 & -2.72 $\pm$ 0.09 & -3.13 $\pm$ 0.08 & +0.34 $\pm$ 0.20 \\
WVK & LMC &  36 & -2.42 $\pm$ 0.18 & -3.29 $\pm$ 0.13 & +0.23 $\pm$ 0.37 \\

 \multicolumn{6}{c}{ \citet{Jacyszyn16} } & OGLE-IV, no reddening correction  \\          
  V & LMC & 2365 & 17.429 $\pm$ 0.004 & -2.672 $\pm$ 0.006 & & all periods    \\
  V & LMC & 2090 & 17.399 $\pm$ 0.005 & -2.629 $\pm$ 0.007 & & $\log P > 0.4$ \\
  V & LMC &  280 & 17.526 $\pm$ 0.010 & -2.964 $\pm$ 0.032 & & $\log P < 0.4$ \\

  V & SMC & 2734  & 17.984 $\pm$ 0.002 & -2.901 $\pm$ 0.005 & & all periods    \\
  V & SMC &  978  & 17.792 $\pm$ 0.006 & -2.648 $\pm$ 0.009 & & $\log P > 0.4$ \\
  V & SMC & 1758  & 18.001 $\pm$ 0.004 & -2.914 $\pm$ 0.015 & & $\log P < 0.4$ \\

 \multicolumn{6}{c}{ \citet{Ngeow09, Ngeow15} } \\                     
  V & LMC & 1675  & 17.115 $\pm$ 0.015 & -2.769 $\pm$ 0.023 & & all periods          \\ 
  V & LMC & 1566  & 17.143 $\pm$ 0.018 & -2.823 $\pm$ 0.031 & & $\log P < 1.0$ \\ 
  V & LMC &  109  & 17.122 $\pm$ 0.195 & -2.746 $\pm$ 0.165 & & $\log P > 1.0$ \\ 

  K & LMC & 1554  & 15.996 $\pm$ 0.010 & -3.194 $\pm$ 0.015 & & all periods \\ 

  V & SMC &  912 & 17.606 $\pm$ 0.028 & -2.660 $\pm$ 0.040 & & all periods     \\ 
  K & SMC &  627 & 16.514 $\pm$ 0.025 & -3.213 $\pm$ 0.032 & & all periods     \\ 

 \multicolumn{6}{c}{ \citet{Ripepi12, Ripepi17} } \\
  K & LMC & 172   & 16.070 $\pm$ 0.017 & -3.295 $\pm$ 0.018 & \\ 
WVK & LMC & 172   & 15.870 $\pm$ 0.013 & -3.325 $\pm$ 0.014 & \\ 

  K & SMC &    & 16.686 $\pm$ 0.009 & -3.513 $\pm$ 0.036 & & $\log P < 0.47$ \\ 
  K & SMC &    & 16.530 $\pm$ 0.018 & -3.224 $\pm$ 0.023 & & $\log P > 0.47$ \\ 

WVK & SMC &   & 16.527 $\pm$ 0.009 & -3.567 $\pm$ 0.034 & & $\log P < 0.47$ \\ 
WVK & SMC &   & 16.375 $\pm$ 0.017 & -3.329 $\pm$ 0.021 & & $\log P > 0.47$ \\ 

 \multicolumn{6}{c}{ \citet{Inno16} } \\ 
  V & LMC & 1526 & 17.172 $\pm$ 0.001 & -2.807 $\pm$ 0.001 & & all periods \\
  K & LMC & 1518 & 16.053 $\pm$ 0.002 & -3.261 $\pm$ 0.003 & & all periods \\
WVK & LMC & 2170 & 15.894 $\pm$ 0.002 & -3.314 $\pm$ 0.002 & & all  periods \\

\hline 

\end{tabular} 
\label{Tab:plfitsLit}
\end{table*}

\subsection{$PLZ$ relation}

\begin{figure}

\centering
\begin{minipage}{0.39\textwidth}
\resizebox{\hsize}{!}{\includegraphics{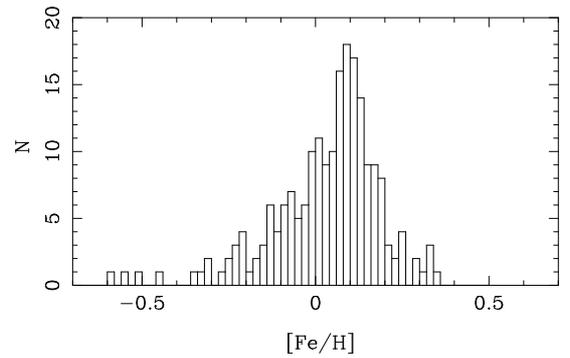}}
\end{minipage}

\caption{Distribution of the [Fe/H] abundance in the reduced sample of 205 objects.
}
\label{Fig:IronD}
\end{figure}

In this section the effect of metallicity is considered.
Figure~\ref{Fig:IronD} shows the histogram over the [Fe/H] abundance of the reduced sample.
The abundance ranges from $-0.59$ to $+0.34$ with an average of $+0.03$ and a median of $+0.06$.
The mean abundance of CCs in the LMC with individually determined accurate metallicities 
(22 stars from \citealt{Luck98,Romaniello08,Lemasle17}) is $-0.35 \pm 0.05$.

Table~\ref{Tab:plzfits} contains the results when fitting the $PLZ$ relation.
The content is similar to Table~\ref{Tab:plfits}; solutions have been derived by either fitting or fixing the slope
of the period dependence, for the two sources of parallax data and for various parallax zero-point offsets.
For the solutions with fixed slope the DM to the LMC is listed in the last column, using the mean abundance and including
the error on the mean abundance for the LMC Cepheids in the error budget.

The parallax zero-point offset again has an important influence on the results.
The zero point and slope of the period dependence change only marginally when also fitting for the metallicity dependence, and the
metallicity dependence itself is absent to marginal (\less 1.5$\sigma$), but there is a degeneracy of the parameters.
Adding the external parallax data makes the metallicity effect stronger.

Solutions 120 and higher are calculated for a fixed period dependence to be able to calculate the DM to the LMC.
The result is a stronger dependence on metallicity (in an absolute sense) with, depending on the parallax zero-point offset and the filter,
a significance that ranges from non-existing to an $\sim 2\sigma$ effect.
For a given parallax zero-point offset the LMC DM is found to be smaller by $\sim 0.03-0.08$ mag than when excluding the metallicity term.

\begin{table*} 

\caption{$PL$ relations of the form $M = \alpha + \beta \log P+ \gamma$ [Fe/H]; 
$N$ is the number of stars in the solution. 
}

\begin{tabular}{rrrrrll} \hline \hline 

 &  $\alpha$   & $\beta$    & $\gamma$ & $N$ & Constraints & LMC DM  \\
 &            &             &         &      &              & (mag) \\  
\hline 

\multicolumn{6}{r}{ FU, $2.7 < P(d) < 35$, $\mid$GOF$\mid<8$, $\epsilon_i<0.001$, $\delta_{\rm PL}$ applied} \\
101 & -1.929 $\pm$ 0.120 & -2.407 $\pm$ 0.143 &  0.326 $\pm$ 0.257 & 194 & V, GDR2, ZPoff= $0$ mas                 \\ 
102 & -1.879 $\pm$ 0.121 & -2.312 $\pm$ 0.142 &  0.117 $\pm$ 0.254 & 194 & V, GDR2, ZPoff= $-0.029$ mas   \\ 
103 & -1.846 $\pm$ 0.120 & -2.259 $\pm$ 0.141 &  0.001 $\pm$ 0.258 & 194 & V, GDR2, ZPoff= $-0.046$ mas   \\ 

104 & -2.917 $\pm$ 0.060 & -3.166 $\pm$ 0.074 &  0.237 $\pm$ 0.171 & 194 & K, GDR2, ZPoff= $0$ mas                   \\ 
105 & -2.865 $\pm$ 0.057 & -3.072 $\pm$ 0.072 &  0.027 $\pm$ 0.164 & 194 & K, GDR2, ZPoff= $-0.029$ mas    \\ 
106 & -2.835 $\pm$ 0.057 & -3.020 $\pm$ 0.071 & -0.090 $\pm$ 0.159 & 194 & K, GDR2, ZPoff= $-0.046$ mas    \\ 

107 & -3.051 $\pm$ 0.055 & -3.266 $\pm$ 0.071 &  0.223 $\pm$ 0.167 & 194 & WVK, GDR2, ZPoff= $0$ mas                   \\ 
108 & -2.999 $\pm$ 0.053 & -3.171 $\pm$ 0.067 &  0.015 $\pm$ 0.161 & 194 & WVK, GDR2, ZPoff= $-0.029$ mas    \\ 
109 & -2.968 $\pm$ 0.052 & -3.120 $\pm$ 0.066 & -0.102 $\pm$ 0.156 & 194 & WVK, GDR2, ZPoff= $-0.046$ mas    \\ 

\\

110 & -1.916 $\pm$ 0.119 & -2.359 $\pm$ 0.136 &  0.106 $\pm$ 0.239 & 205 & V, GDR2+Ext, ZPoff= $0$ mas                   \\ 
111 & -1.834 $\pm$ 0.118 & -2.250 $\pm$ 0.136 & -0.064 $\pm$ 0.236 & 205 & V, GDR2+Ext, ZPoff= $-0.046$ mas    \\ 
112 & -2.907 $\pm$ 0.058 & -3.109 $\pm$ 0.069 & -0.003 $\pm$ 0.163 & 205 & K, GDR2+Ext , ZPoff= $0$ mas                  \\ 
113 & -2.824 $\pm$ 0.057 & -2.999 $\pm$ 0.067 & -0.189 $\pm$ 0.139 & 205 & K, GDR2+Ext, ZPoff= $-0.046$ mas    \\ 
114 & -3.041 $\pm$ 0.053 & -3.206 $\pm$ 0.065 & -0.013 $\pm$ 0.160 & 205 & WVK, GDR2+Ext, ZPoff= $0$ mas                 \\ 
115 & -2.976 $\pm$ 0.052 & -3.096 $\pm$ 0.062 & -0.204 $\pm$ 0.140 & 205 & WVK, GDR2+Ext, ZPoff= $-0.046$ mas  \\ 

\\

120 & -1.617 $\pm$ 0.033 & -2.810 fixed & 0.388 $\pm$ 0.271 & 194 & V, GDR2, ZPoff= $0$ mas        & 18.925 $\pm$ 0.102  \\ 
121 & -1.494 $\pm$ 0.033 & -2.810 fixed & 0.197 $\pm$ 0.278 & 194 & V, GDR2, ZPoff= $-0.029$ mas  & 18.735 $\pm$ 0.103  \\ 
122 & -1.424 $\pm$ 0.034 & -2.810 fixed & 0.095 $\pm$ 0.284 & 194 & V, GDR2, ZPoff= $-0.046$ mas  & 18.629 $\pm$ 0.105 \\ 

125 & -2.846 $\pm$ 0.019 & -3.260 fixed &  0.252 $\pm$ 0.166 & 194 & K, GDR2, ZPoff= $0$ mas         & 18.987 $\pm$ 0.062  \\ 
126 & -2.721 $\pm$ 0.017 & -3.260 fixed &  0.054 $\pm$ 0.162 & 194 & K, GDR2, ZPoff= $-0.029$ mas  & 18.792 $\pm$ 0.059  \\ 
127 & -2.651 $\pm$ 0.017 & -3.260 fixed & -0.053 $\pm$ 0.161 & 194 & K, GDR2, ZPoff= $-0.046$ mas  & 18.685 $\pm$ 0.059  \\ 

130 & -3.006 $\pm$ 0.018 & -3.325 fixed &  0.231 $\pm$ 0.166  & 194 & WVK, GDR2, ZPoff= $0$ mas       & 18.957 $\pm$ 0.063 \\ 
131 & -2.881 $\pm$ 0.016 & -3.325 fixed &  0.038 $\pm$ 0.156  & 194 & WVK, GDR2, ZPoff= $-0.029$ mas  & 18.764 $\pm$ 0.058   \\ 
132 & -2.811 $\pm$ 0.016 & -3.325 fixed & -0.070 $\pm$ 0.157  & 194 & WVK, GDR2, ZPoff= $-0.046$ mas  & 18.656  $\pm$ 0.059  \\ 

\\

140 & -1.566 $\pm$ 0.034 & -2.810 fixed &  0.136 $\pm$ 0.250 & 205 & V, GDR2+Ext, ZPoff= $0$ mas           & 18.786 $\pm$ 0.094 \\ 
141 & -1.460 $\pm$ 0.034 & -2.810 fixed &  0.020 $\pm$ 0.254 & 205 & V, GDR2+Ext, ZPoff= $-0.029$ mas & 18.639 $\pm$ 0.095  \\ 
142 & -1.400 $\pm$ 0.033 & -2.810 fixed & -0.041 $\pm$ 0.260 & 205 & V, GDR2+Ext, ZPoff= $-0.046$ mas & 18.558 $\pm$ 0.097 \\ 

145 & -2.789 $\pm$ 0.017 & -3.260 fixed &  0.013 $\pm$ 0.163 & 205 & K, GDR2+Ext, ZPoff= $0$ mas        & 18.847 $\pm$ 0.060  \\ 
146 & -2.681$\pm$  0.016 & -3.260 fixed & -0.106 $\pm$ 0.148 & 205 & K, GDR2+Ext, ZPoff= $-0.029$ mas & 18.697 $\pm$ 0.055   \\ 
147 & -2.620 $\pm$ 0.016 & -3.260 fixed & -0.168 $\pm$ 0.146 & 205 & K, GDR2+Ext, ZPoff= $-0.046$ mas & 18.614 $\pm$ 0.054  \\ 
148 & -2.599 $\pm$ 0.016 & -3.260 fixed & -0.193 $\pm$ 0.143 & 205 & K, GDR2+Ext, ZPoff= $-0.052$ mas & 18.584 $\pm$ 0.053  \\ 

150 & -2.948 $\pm$ 0.017 & -3.325 fixed & -0.002 $\pm$ 0.156  & 205 & WVK, GDR2+Ext, ZPoff= $0$ mas         & 18.817 $\pm$ 0.058 \\ 
151 & -2.840 $\pm$ 0.015 & -3.325 fixed & -0.125 $\pm$ 0.146  & 205 & WVK, GDR2+Ext, ZPoff= $-0.029$ mas & 18.666 $\pm$ 0.055 \\ 
152 & -2.779 $\pm$ 0.015 & -3.325 fixed & -0.188 $\pm$ 0.142  & 205 & WVK, GDR2+Ext, ZPoff= $-0.046$ mas & 18.583 $\pm$ 0.058 \\ 
153 & -2.758 $\pm$ 0.015 & -3.325 fixed & -0.211 $\pm$ 0.141  & 205 & WVK, GDR2+Ext, ZPoff= $-0.052$ mas & 18.554 $\pm$ 0.054 \\ 
154 & -2.713 $\pm$ 0.015 & -3.325 fixed & -0.260 $\pm$ 0.140  & 205 & WVK, GDR2+Ext, ZPoff= $-0.065$ mas & 18.492 $\pm$ 0.054 \\ 

\hline

\end{tabular} 
\label{Tab:plzfits}
\end{table*}

\begin{table*} 

\caption{$PL$ relations of the form $M = \alpha + \beta \log P+ \gamma$ [Fe/H] and
variation in the parameters; $N$ is the number of stars in the solution. 
 }

\begin{tabular}{rrrrrll} \hline \hline 

 &  $\alpha$   & $\beta$    & $\gamma$ & $N$ & Constraints & LMC DM  \\  
 &            &             &         &      &               & (mag) \\  
\hline 

\multicolumn{6}{r}{Standard: FU, $2.7 < P(d) < 35$, $\mid$GOF$\mid<8$, $\epsilon_i<0.001$, $\delta_{\rm PL}$ applied, GDR2+Ext set, ZPoff= $-0.049$ mas} \\

160 & $-1.390 \pm$ 0.033 & $-2.810$ fixed & $-0.051 \pm$ 0.260 & 205 & V, standard                      & 18.544 $\pm$ 0.097   \\ 
161 & $-1.470 \pm$ 0.060 & $-2.810$ fixed & $-0.042 \pm$ 0.453 & 205 & V, $\delta_{\rm PL} \cdot 2 = 0.40$ mag & 18.627 $\pm$ 0.170  \\ 
162 & $-1.372 \pm$ 0.035 & $-2.810$ fixed & $-0.149 \pm$ 0.280 & 205 & V, $\spi \cdot 1.5$ for GDR2       & 18.492 $\pm$ 0.104   \\ 
163 & $-1.442 \pm$ 0.032 & $-2.810$ fixed & $-0.084 \pm$ 0.248 & 205 & V, $E(B-V)\cdot 1.05$              & 18.585 $\pm$ 0.093   \\ 
164 & $-1.394 \pm$ 0.039 & $-2.810$ fixed & $-0.005 \pm$ 0.264 & 205 & V, other pick of Iron abundance    & 18.564 $\pm$ 0.100   \\ 
165 & $-1.388 \pm$ 0.033 & $-2.810$ fixed & $-0.060 \pm$ 0.268 & 205 & V, $\sigma_{\rm [Fe/H]} = 0.04$ dex      & 18.539 $\pm$ 0.099   \\ 
166 & $-1.405 \pm$ 0.053 & $-2.810$ fixed & $-0.230 \pm$ 0.443 &  28 & V, $\spi/\pi < 0.05$               & 18.497 $\pm$ 0.164   \\ 
167 & $-1.389 \pm$ 0.033 & $-2.810$ fixed & $+0.010 \pm$ 0.262 & 208 & V, $P >2.7$d                       & 18.565 $\pm$ 0.097   \\ 
168 & $-1.385 \pm$ 0.033 & $-2.810$ fixed & $-0.047 \pm$ 0.243 & 217 & V, $P >2.0$d                       & 18.541 $\pm$ 0.091   \\ 

\\

170 & $-2.610 \pm$ 0.016 & $-3.260$ fixed & $-0.182 \pm$ 0.145 & 205 & K, standard                     & 18.599 $\pm$ 0.054 \\
171 & $-2.618 \pm$ 0.023 & $-3.260$ fixed & $-0.183 \pm$ 0.177 & 205 & K, $\delta_{\rm PL} \cdot 2 = 0.13$ mag & 18.607 $\pm$ 0.067 \\ 
172 & $-2.584 \pm$ 0.019 & $-3.260$ fixed & $-0.279 \pm$ 0.164 & 205 & K, $\spi \cdot 1.5$ for GDR2       & 18.539 $\pm$ 0.062 \\ 
173 & $-2.614 \pm$ 0.016 & $-3.260$ fixed & $-0.185 \pm$ 0.146 & 205 & K, $E(B-V) \cdot 1.05$              & 18.602 $\pm$ 0.054 \\ 
174 & $-2.591 \pm$ 0.016 & $-3.260$ fixed & $-0.171 \pm$ 0.143 & 205 & K, $A_{K}/A_{V}= 0.10$        & 18.584 $\pm$ 0.053 \\ 
175 & $-2.611 \pm$ 0.019 & $-3.260$ fixed & $-0.113 \pm$ 0.157 & 205 & K, other pick of Iron abundance    & 18.624 $\pm$ 0.058 \\ 
176 & $-2.604 \pm$ 0.016 & $-3.260$ fixed & $-0.259 \pm$ 0.140 & 205 & K, $\sigma_{\rm [Fe/H]} = 0.04$ dex      & 18.566 $\pm$ 0.053 \\ 
177 & $-2.602 \pm$ 0.025 & $-3.260$ fixed & $-0.267 \pm$ 0.260 &  28 & K, $\spi/\pi < 0.05$               & 18.562 $\pm$ 0.095 \\ 
178 & $-2.609 \pm$ 0.017 & $-3.260$ fixed & $-0.141 \pm$ 0.151 & 208 & K, $P >2.7$d                       & 18.612 $\pm$ 0.056 \\ 
179 & $-2.605 \pm$ 0.016 & $-3.260$ fixed & $-0.207 \pm$ 0.146 & 217 & K, $P >2.0$d                       & 18.586 $\pm$ 0.055 \\ 

\\

180 & $-2.769 \pm$ 0.015 & $-3.325$ fixed & $-0.200 \pm$ 0.141  & 205 & WVK, standard                     & 18.569 $\pm$ 0.054 \\
181 & $-2.773 \pm$ 0.019 & $-3.325$ fixed & $-0.197 \pm$ 0.157  & 205 & WVK, $\delta_{\rm PL} \cdot 2 = 0.09$ mag & 18.574 $\pm$ 0.060 \\ 
182 & $-2.741 \pm$ 0.018 & $-3.325$ fixed & $-0.297 \pm$ 0.166  & 205 & WVK, $\spi \cdot 1.5$ for GDR2      & 18.507 $\pm$ 0.064 \\ 
183 & $-2.769 \pm$ 0.015 & $-3.325$ fixed & $-0.200 \pm$ 0.141  & 205 & WVK, $E(B-V) \cdot 1.05$             & 18.569 $\pm$ 0.054 \\ 
184 & $-2.748 \pm$ 0.015 & $-3.325$ fixed & $-0.186 \pm$ 0.141  & 205 & WVK, $A_{K}/A_{V}= 0.10$       & 18.553 $\pm$ 0.054 \\ 
185 & $-2.770 \pm$ 0.019 & $-3.325$ fixed & $-0.126 \pm$ 0.155  & 205 & WVK, other pick of Iron abundance   & 18.596 $\pm$ 0.059 \\ 
186 & $-2.762 \pm$ 0.015 & $-3.325$ fixed & $-0.285 \pm$ 0.133  & 205 & WVK, $\sigma_{\rm [Fe/H]} = 0.04$ dex   & 18.532 $\pm$ 0.053 \\ 
187 & $-2.768 \pm$ 0.015 & $-3.325$ fixed & $-0.199 \pm$ 0.143  & 205 & WVK, $\sigma_{V} \cdot 2$, $\sigma_{K} \cdot 2$  & 18.568 $\pm$ 0.055 \\ 
188 & $-2.784 \pm$ 0.015 & $-3.325$ fixed & $-0.191 \pm$ 0.145  & 205 & WVK, No NIR transformation  & 18.587 $\pm$ 0.055 \\ 
189 & $-2.756 \pm$ 0.024 & $-3.325$ fixed & $-0.273 \pm$ 0.247  &  28 & WVK, $\spi/\pi < 0.05$              & 18.530 $\pm$ 0.092 \\ 
190 & $-2.768 \pm$ 0.016 & $-3.325$ fixed & $-0.158 \pm$ 0.147  & 208 & WVK, $P >2.7$d                      & 18.583 $\pm$ 0.056 \\ 
191 & $-2.763 \pm$ 0.015 & $-3.325$ fixed & $-0.224 \pm$ 0.140  & 217 & WVK, $P >2.0$d                      & 18.555 $\pm$ 0.054 \\ 

\\
200 & $-1.840 \pm$ 0.118 & $-2.243 \pm$ 0.137 & 0.0 fixed  & 205 & V, standard                      &  \\ 
201 & $-2.827 \pm$ 0.055 & $-3.000 \pm$ 0.065 & 0.0 fixed  & 205 & K, standard                      &  \\ 
202 & $-2.961 \pm$ 0.051 & $-3.098 \pm$ 0.060 & 0.0 fixed  & 205 & WVK, standard                    &  \\ 

\hline

\end{tabular} 
\label{Tab:plzfitsfinal}
\end{table*}

\section{Discussion and summary}
\label{S-Dis}

From  an initial sample of 452 Galactic Cepheids with accurate [Fe/H] abundances, period-luminosity and period-luminosity-metallicity relations 
have been derived based on parallax data from \G\ DR2, supplemented with accurate non-\G\ parallax data when available, 
for a final sample of about 200 FU mode Cepheids with good astrometric solutions.

The influence of a parallax zero-point offset on the derived $PL(Z)$ relation is large, which means  that the 
current GDR2 results do not allow us to improve on the existing calibration of the relation or on the distance to the LMC.
The zero point, the slope of the period dependence,  and any metallicity dependence of the $PL(Z)$ relations 
are correlated with any assumed parallax zero-point offset.

Based on a comparison for nine CCs with the best non-\G\ parallaxes (mostly from {\it HST} data) a 
parallax zero-point offset of $-0.049 \pm 0.018$ mas,  consistent with other values that appeared in the literature after the release of GDR2,
is derived  from RGB stars using {\it Kepler} and {\it APOGEE} data  (about $-0.053$ mas, \citealt{Zinn18}), 
eclipsing binaries ($-0.082 \pm 0.033$ mas, \citealt{Stassun18}),
a sample of 50 CCs ($-0.046 \pm 0.013$ mas, \citealt{RiessGDR2}), and
RR Lyrae stars ($\sim -0.056$ mas, \citealt{Muraveva18}).

For a parallax zero-point offset of $-0.049$ mas a final list of calculations, investigating the influence of some other 
parameters is given Table~\ref{Tab:plzfitsfinal}. The slope of the period dependence has been fixed (solutions 160--190),  
and the resulting  DM to the LMC is listed in the last column. Variations in the assumed dispersion  in the 
photometry, reddening, or period cuts 
have a relatively small impact on the results (especially in $K$ and $WVK$). 
These models address some concerns that might arise when combining many sources of photometry, reddening, and iron abundances from the literature. 
Increasing the error in the mean $V$ and $K$ magnitude, or not applying a transformation of the different NIR magnitude systems at all has very little impact.

The most noticeable effect is when the parallax errors in GDR2 are underestimated as this changes the weight of the 
GDR2 sample with respect to the stars with a non-\G\ parallax. \citet{Lindegren18} in their Appendix~A hint at 
the fact that the errors in the five astrometric parameters may still be underestimated even after correcting for the DOF bug.
This issue will likely be resolved in future releases. The next most important effect is the iron abundance. 
A smaller error in its determination would help, but equally important is a homogeneous metallicity scale (see \citealt{Proxauf18} for new efforts in this direction).

\begin{figure}
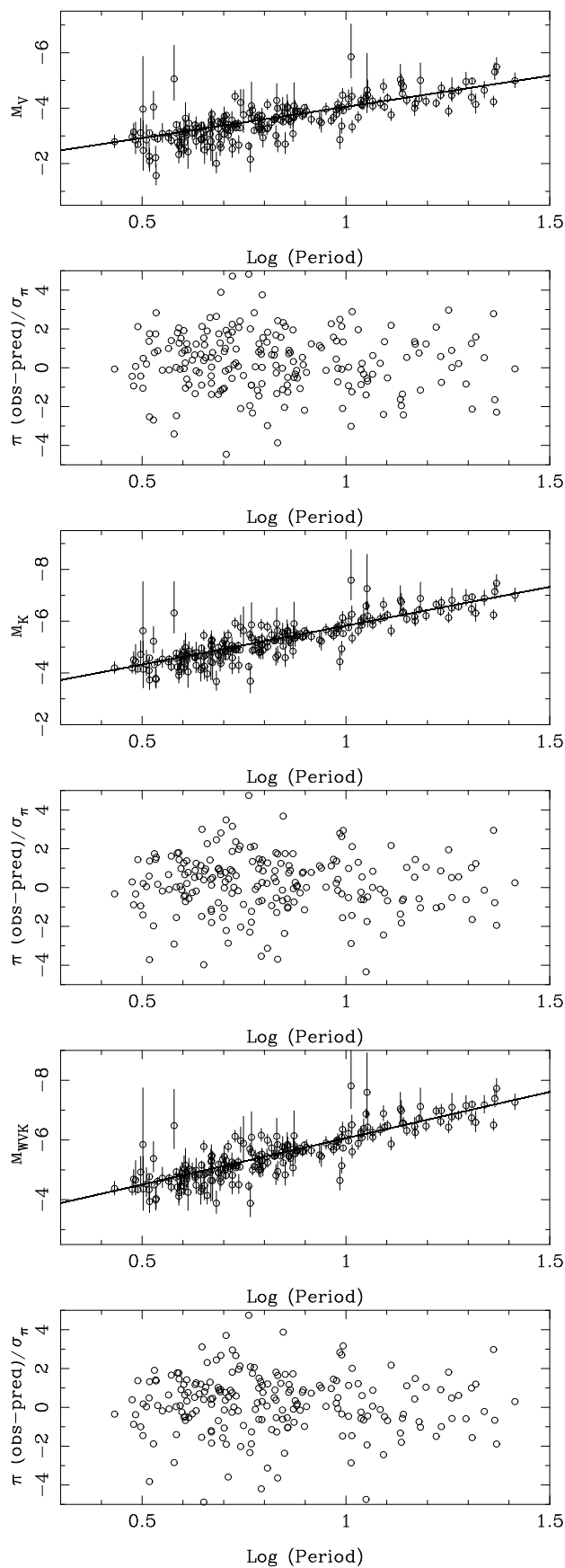


\centering

\begin{minipage}{0.44\textwidth}
\resizebox{\hsize}{!}{\includegraphics{MlogP_V.ps}}
\end{minipage}
\begin{minipage}{0.44\textwidth}
\resizebox{\hsize}{!}{\includegraphics{par_parpred_V.ps}}
\end{minipage}

\begin{minipage}{0.44\textwidth}
\resizebox{\hsize}{!}{\includegraphics{MlogP_K.ps}}
\end{minipage}
\begin{minipage}{0.44\textwidth}
\resizebox{\hsize}{!}{\includegraphics{par_parpred_K.ps}}
\end{minipage}

\begin{minipage}{0.44\textwidth}
\resizebox{\hsize}{!}{\includegraphics{MlogP_WVK.ps}}
\end{minipage}
\begin{minipage}{0.44\textwidth}
\resizebox{\hsize}{!}{\includegraphics{par_parpred_WVK.ps}}
\end{minipage}

\caption{$PL$ relations in the $V$, $K$, and $WVK$ bands (solutions 200--202 from Table~\ref{Tab:plzfitsfinal}).
For each filter the second panel gives the deviation between observed and predicted parallax (limited to $\pm5~\sigma$).
}
\label{Fig:PL}
\end{figure}

Solutions 200--202 give the current best estimate of the $PL$ relation (for the assumed parallax zero-point offset), 
without a metallicity dependence as the current data and analysis does not allow us to prove or disprove  a dependence.
Figure~\ref{Fig:PL} shows these relations in the three bands considered. No additional sigma-clipping in magnitude space has been applied 
(as is common in deriving $PL$ relations in the LMC). Only stars that were systematic outliers in $\sigma$-space have been removed, 
as explained in Sect. \ref{S-Res}.
As noted before, the slopes of the galactic $PL$ relation are shallower than those derived in the LMC by $\sim 3\sigma$ (for the assumed parallax zero-point offset).
If the slope is fixed to values that have been derived for LMC Cepheids, the derived parallax zero-point offset suggests a LMC DM that is larger than commonly adopted.
Conversely, a LMC DM of around 18.50 requires a zero-point offset closer to $-0.1$~mas.

The results in this paper show that the parallax zero-point offset should be known to a level of $\less3~\mu$as to have a $\less$0.01 mag effect 
on zero point of the $PL$ relation and the DM to the LMC.
It will remain important to have accurate non-\G\ based parallaxes as a control sample, like the ongoing programme using {\it HST} \citep{Riess14,Casertano16,Riess18}.

\begin{acknowledgements}
This work has made use of data from the European Space Agency (ESA) mission {\it Gaia} 
(\url{http://www.cosmos.esa.int/gaia}), processed by the {\it Gaia} Data Processing and Analysis Consortium 
(DPAC, \url{http://www.cosmos.esa.int/web/gaia/dpac/consortium}). 
Funding for the DPAC has been provided by national institutions, in particular
the institutions participating in the {\it Gaia} Multilateral Agreement.
This research has made use of the SIMBAD database and the VizieR catalogue access tool 
operated at CDS, Strasbourg, France.
The original description of the VizieR service was published in A\&AS 143, 23.
\end{acknowledgements}

 \bibliographystyle{aa.bst}
        \bibliography{references.bib}

\begin{appendix}

\section{Initial run of models}
\label{AppIni}

Table~\ref{Tab:App1} shows the results of fitting period-luminosity relations in the $V$, $K$, and $WVK$ bands based on several
selection criteria, which include the goodness of fit, fundamental or overtone pulsators, period cuts, and parallax zero-point offsets.
This initial set of models was calculated to investigate the parameter space and arrive at a `best' sample, which  is used in the main text.

The table is organised as follows. The models are grouped by the parallax that was used, 
GDR2-only or GDR2+External values, as outlined in Sect. \ref{S-Res}, and by filter ($V$,  $K$, or $WVK$). Then the solutions
are given applying the various constraints listed in   Col. 4.

Inspecting the results shows that the overtone pulsators (using their fundamentalised period) have significantly different
$PL$ relations than the FU mode pulsators. This was not investigated further, but led to the conservative choice of using only FU mode
Cepheids in the final sample. 

A cut on period was investigated and finally implemented (period between 2.7 and 35 days). 
The exact values are less important, but they were used to avoid a possible contamination by unrecognised overtone pulsators 
(at the short period end) and to make a fairer comparison to samples in the Magellanic Clouds where such period cuts are typically used.
 
Stars for the final sample were selected to have  $\mid$GOF$\mid<8$ and no excess noise.
This is related to the problem of very poor astrometric solutions, which can occur for the brightest stars, but also due
to binarity (see main text).

\section{ The Polaris system }
\label{AppPol}

In this appendix we discuss the distance to the Polaris system, the CC $\alpha$ UMi (Polaris~A\footnote{Or Aa to be precise, 
as this is a binary system  itself, with a companion that is of order 5 mag fainter at less than 0.2\arcsec\ distance \citep{Evans2008}.}) and 
its companion Polaris~B, located at about 18\arcsec. 
Polaris~A is listed in GDR2 with the largest GOF of the CCs in the sample and without parallax.
The parallax used is the average of the available \Hp\ parallaxes (Table~\ref{Tab:Hipp}), but as FO pulsators are excluded in the final analysis, 
Polaris~A is not actually used in this paper.
Polaris~B is listed in GDR2 with a parallax of 7.292 $\pm$ 0.028 mas. The astrometric solution has zero excess noise and a GOF value of about 12, which 
means it would not make it through the selection criterion of GOF $<8$ that was applied to the CCs in the sample.

Polaris B has a recent {\it HST} based parallax of 6.26 $\pm$ 0.26 mas \citep{Bond18} which differs by 4$\sigma$ from
the GDR2 value.
\citet{Anderson18} showed that adopting the parallax by \citet{Bond18} for Polaris~A a consistent picture emerged of Polaris
as a 7~\msol\ first-overtone CC near the hot boundary of the first IS crossing.
This conclusion was based on the Geneva set of rotating stellar evolutionary models \citep{Anderson16} and considering
various constraints (the rate of period change, the Wesenheit $WVI$ $PL$ relation, the location in the $M_V - (B-V)$ colour-magnitude diagram, 
the interferometrically determined radius, spectroscopic N/C and N/O abundance ratios, and a dynamical mass measurement).
\citet{Anderson18} does not discuss the quality of the  model that can be obtained using the larger \Hp\ parallax\footnote{His paper appeared pre-GDR2.}.

Based on the current best estimates of the $PL$ relations (solutions 200--202 in Table~\ref{Tab:plzfitsfinal})
the photometric parallax of Polaris A is predicted using Monte Carlo simulations, that is, taking into
account the uncertainty in $V$ and $K$ photometry and reddening value, and the width of the instability strip. 
A fundamentalised period of 5.673 days is used.
The results are listed in Table~\ref{Tab:Pol}.

\begin{figure}
\centering

\begin{minipage}{0.44\textwidth}
\resizebox{\hsize}{!}{\includegraphics{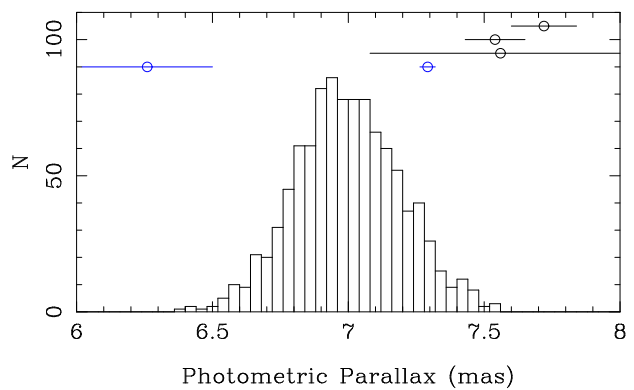}}
\end{minipage}

\caption{Distribution of the photometric parallax predicted for Polaris based on the $WVK$ $PL$ relation.
The blue lines give the parallax and error determined for Polaris B (6.3 mas from \citealt{Bond18}, 7.3 mas from GDR2) and 
the black lines the various estimates from \Hp\ for Polaris A (see Table~\ref{Tab:Hipp}).
}
\label{Fig:Polaris}
\end{figure}

For the standard case the predicted parallax distribution based on the $WVK$ relation gives a median and 
dispersion (based on 1.483 $\times$ the median absolute deviation, which is equivalent to $\sigma$ in the case of a Gaussian distribution) of 6.99 $\pm$ 0.19 mas.
The $K$ band gives a consistent result of 7.12 $\pm$ 0.23 mas; the $V$ band also gives a  consistent, but more uncertain, value of 8.1 $\pm$ 0.8 mas.
Figure~\ref{Fig:Polaris} shows the distribution in predicted parallax based on the $WVK$ $PL$ relation and compares it to the various estimates in the literature.
These parallax estimates depend on the assumed parallax zero-point offset, and Table~\ref{Tab:Pol} list values for some other offsets for the Wesenheit $WVK$ magnitude only, as it gives the best precision.

The calculations and the resulting parallax distribution have assumed so  far that Polaris~A is located in the centre of the IS with the width of the IS 
represented by a Gaussian with width $\delta_{\rm PL}= 0.049$ mag. 
\citet{Anderson18} have argued that Polaris~A is located at the blue edge (i.e. the hotter end) of the IS (based on the shorter parallax by \citealt{Bond18}).
Adding +0.1 mag to the $WVK$ colour of Polaris A to simulate the colour it would have at the centre of the IS results in a 
parallax estimate of 6.69 $\pm$ 0.10 mas (assuming $\delta_{\rm PL}= 0$ mag now). 

A particular problem for Polaris~A is that its $K$-band magnitude is not very well established, and that interferometric observations have revealed
a small excess of 0.016 $\pm$ 0.004 mag in $K$ \citep{Merand06} due to the presence of an extended circumstellar envelope. 
Making the standard adopted value (see the main text) of $K= 0.652 \pm 0.028$ mag fainter by this amount (and adding the error in quadrature) has a small effect.
The actual value has a potentially larger impact. 
\cite{vanL07} also use the {\it COBE-DIRBE} flux of Polaris, but adopting a different calibration scheme and magnitudes for Procyon 
arrive at $J$ = 0.98 mag and $K$ = 0.60 mag on the SAAO system (with no error bars given).
Using the transformation formulae in \citet{Carpenter2001}, we arrive at $K= 0.58$ mag on the 2MASS system.
Using this value (with the same error bar of 0.028 mag as we derived) results in a significantly larger parallax.

In summary, the predicted photometric parallax is not conclusive. It is roughly halfway between the \Hp\ parallax estimates 
for Polaris~A and the GDR2 estimate for Polaris~B, and the estimate by \citet{Bond18}.

\begin{table*}
 \caption{Photometric parallax predictions for Polaris.}
\centering
\small
  \begin{tabular}{cl}
  \hline
parallax     &  Remarks  \\ 
(mas)        &       \\ 
  \hline

6.989 $\pm$ 0.186 & WVK (Sol.~200 from Tab.~\ref{Tab:plzfitsfinal}) \\
7.122 $\pm$ 0.234 & K (Sol.~201) \\
8.080 $\pm$ 0.759 & V (Sol.~202) \\
\\
6.484 $\pm$ 0.174 & WVK, ZPoff= $0.0$ mas \\
6.782 $\pm$ 0.181 & WVK, ZPoff= $-0.029$ mas \\
6.989 $\pm$ 0.186 & WVK, ZPoff= $-0.049$ mas (Sol.~200) \\
7.103 $\pm$ 0.189 & WVK, ZPoff= $-0.060$ mas \\ 
7.310 $\pm$ 0.194 & WVK, ZPoff= $-0.080$ mas \\ %
7.515 $\pm$ 0.198 & WVK, ZPoff= $-0.100$ mas \\ %
\\
6.687 $\pm$ 0.102 & WVK, ZPoff= $-0.049$ mas, location in the IS, $WVK+0.1$ mag, $\delta_{\rm PL}=0$ mag \\
6.931 $\pm$ 0.185 & WVK, ZPoff= $-0.049$ mas, NIR excess, $\Delta K$= 0.016 $\pm$ 0.004 mag \\
7.256 $\pm$ 0.194 & WVK, ZPoff= $-0.049$ mas, alternative $K$ mag, $K= 0.58 \pm 0.028$ mag \\

\hline
\end{tabular}

\label{Tab:Pol}
\end{table*}

\end{appendix}

\setcounter{table}{0}
\longtab[1]{
\begin{landscape}
\begin{longtable}{rlrrcrrrcccrrr}
  \caption{\label{Tab:Targets} Sample of stars. [TABEL IS AVAILABLE AT THE CDS]} \\
  \hline  \hline
Name      & Type     &   Period &  $V$  & Ref &  $J$  &   $H$  &   $K$  &  Ref &    $E(B-V)$         & Ref &  [Fe/H]$_1$  & [Fe/H]$_2$ &  [Fe/H]$_3$   \\ 
          &          &    (d)   &   (mag)   &     &      (mag)   &   (mag)    &   (mag)   &  &   (mag)        &     &           &         &            \\ 
\hline
\endfirsthead
\caption{continued.}\\
\hline\hline
Name      & Type     &  Period &  Vmag  & Ref &  $J$ &   $H$  &   $K$ &  Ref &    $E(B-V)$         & Ref &  [Fe/H]$_1$  & [Fe/H]$_2$  & [Fe/H]$_3$   \\ 
          &          &    (d)   &   (mag)   &     &    (mag)  &     (mag)  &    (mag)  &      &      (mag)          &     &          &                     \\ 
\hline
\endhead
\hline
\endfoot
AA    Gem  &   DCEP       & 11.302 &  9.720 &  1 &  7.646 &  7.189 &  7.086 &    (1)               & 0.345 $\pm$ 0.036 &   (1)  &          -0.08 2   &           -0.14 3  &           -0.22 5 \\  
AA    Mon  &   DCEP       &  3.938 & 12.740 &  1 &  9.729 &  9.186 &  8.989 &    (1)               & 0.765 $\pm$ 0.017 &   (1)  &          -0.12 1   &           -0.09 3  &           -0.15 5 \\  
AB    Cam  &   DCEP       &  5.788 & 11.860 &  1 &  9.418 &  8.936 &  8.746 &    (7)               & 0.622 $\pm$ 0.035 &   (1)  &          -0.11 1   &           -0.07 4  &           -0.03 5 \\  
AC    Mon  &   DCEP       &  8.014 & 10.100 &  1 &  7.579 &  7.054 &  6.885 &    (1)               & 0.507 $\pm$ 0.033 &   (1)  &          -0.06 1   &           -0.03 3  &           -0.17 5 \\  
AD    Cam  &   DCEP       & 11.261 & 12.560 &  1 &  8.988 &  8.363 &  8.109 &    (7)               & 0.873 $\pm$ 0.012 &   (1)  &          -0.28 1   &           -0.25 4  &           -0.17 5 \\  
AD    Cru  &   DCEP       &  6.398 & 11.060 &  1 &  8.198 &  7.661 &  7.387 &    (7)               & 0.640 $\pm$ 0.012 &   (1)  &           0.08 1   &            0.11 3  &            0.11 5 \\  
AD    Gem  &   DCEP       &  3.788 &  9.850 &  1 &  8.468 &  8.140 &  8.066 &    (1)               & 0.206 $\pm$ 0.048 &   (1)  &          -0.14 1   &           -0.15 3  &           -0.11 5 \\  
AD    Pup  &   DCEP       & 13.596 &  9.920 &  1 &  7.723 &  7.341 &  7.144 &    (7)               & 0.363 $\pm$ 0.020 &   (1)  &          -0.06 1   &           -0.03 3  &           -0.12 5 \\  
AE    Tau  &   DCEP       &  3.897 & 11.700 &  1 &  9.442 &  8.949 &  8.750 &    (7)               & 0.568 $\pm$ 0.056 &   (1)  &          -0.21 1   &           -0.18 3  &           -0.14 5 \\  
AE    Vel  &   DCEP       &  7.134 & 10.240 &  1 &  7.589 &  7.096 &  6.867 &    (7)               & 0.691 $\pm$ 0.055 &   (1)  &           0.11 1   &            0.14 3  &            0.11 5 \\  
AG    Cru  &   DCEP       &  3.837 &  8.220 &  1 &  6.629 &  6.368 &  6.200 &    (6)(10)   =  5    & 0.242 $\pm$ 0.020 &   (1)  &           0.05 1   &            0.08 3  &           -0.08 5 \\  
AH    Vel  &   DCEPS      &  4.227 &  5.690 &  1 &  4.568 &  4.327 &  4.235 &    (3)               & 0.036 $\pm$ 0.019 &   (1)  &           0.09 1   &            0.19 3  &            0.11 5 \\  
alf   UMi  &   DCEPS      &  3.970 &  1.980 &  1 &  0.941 &  0.460 &  0.652 &  (13)                & 0.016 $\pm$ 0.009 &   (1)  &                    &                    &            0.15 5 \\  
AN    Aur  &   DCEP       & 10.291 & 10.450 &  1 &  7.932 &  7.434 &  7.274 &    (1)               & 0.542 $\pm$ 0.040 &   (1)  &          -0.13 1   &           -0.10 3  &           -0.10 5 \\  
AO    Aur  &   DCEP       &  6.763 & 10.850 &  1 &  8.640 &  8.188 &  8.064 &    (1)               & 0.437 $\pm$ 0.044 &   (1)  &          -0.27 1   &           -0.26 4  &           -0.21 5 \\  
AO    CMa  &   DCEP       &  5.816 & 12.070 &  1 &  9.154 &  8.595 &  8.366 &    (3)               & 0.694 $\pm$ 0.069 &   (1)  &           0.01 1   &                    &           -0.09 5 \\  
AP    Pup  &   DCEP       &  5.084 &  7.380 &  1 &  5.788 &  5.443 &  5.322 &    (3)               & 0.250 $\pm$ 0.034 &   (1)  &          -0.02 1   &            0.08 3  &            0.06 5 \\  
AP    Sgr  &   DCEP       &  5.058 &  6.950 &  1 &  5.350 &  4.970 &  4.890 &    (6)               & 0.184 $\pm$ 0.015 &   (1)  &           0.16 1   &            0.17 3  &            0.15 5 \\  
AQ    Car  &   DCEP       &  9.769 &  8.850 &  1 &  7.164 &  6.811 &  6.573 &    (10)(11)  =  2    & 0.168 $\pm$ 0.013 &   (1)  &          -0.16 1   &            0.03 3  &            0.06 5 \\  
AQ    Pup  &   DCEP       & 30.149 &  8.710 &  1 &  6.099 &  5.481 &  5.297 &    (2)               & 0.531 $\pm$ 0.017 &   (1)  &           0.06 1   &            0.04 3  &           -0.03 5 \\  
AS    Per  &   DCEP       &  4.973 &  9.720 &  1 &  6.934 &  6.465 &  6.295 &    (1)               & 0.684 $\pm$ 0.041 &   (1)  &           0.11 1   &            0.14 3  &            0.15 5 \\  
AT    Pup  &   DCEP       &  6.665 &  7.980 &  1 &  6.289 &  5.963 &  5.866 &    (10)              & 0.166 $\pm$ 0.011 &   (1)  &          -0.08 1   &            0.05 3  &           -0.09 5 \\  
AV    Cir  &   DCEPS      &  3.065 &  7.410 &  1 &  5.582 &  5.236 &  5.093 &    (3)               & 0.374 $\pm$ 0.011 &   (1)  &           0.14 1   &            0.17 3  &            0.15 5 \\  
AV    Sgr  &   DCEP       & 15.415 & 11.290 &  1 &  6.878 &  6.089 &  5.730 &    (3)               & 1.238 $\pm$ 0.027 &   (1)  &           0.35 1   &            0.41 3  &            0.40 5 \\  

\end{longtable}
\tablefoot{
Column~1. The variable star name. The last three objects refer to ASAS~J181024-2049.6,  ASAS~J171305-4323.0, and
ASAS~J184443-0401.5 (or BD~-04~4569), respectively.
Column~2. The classification of the variability, mainly from the VSX catalogue \citep{Watson06}. See the main text for discussion.
A `?' is added when there are different classifications in the literature. 
Based on the analysis some classifications are changed, see main text. These stars are marked by a $\dagger$.
Column~3. The pulsation period, taken from the VSX catalogue \citep{Watson06}.
Columns~4-5. $V$-band magnitude and reference: 1= \citet{Melnik15}, 2= \citet{Berdnikov2000},
3= \citet{MB1985}, 4= \citet{Gr2013}, 5= \citet{MarAnd15}, 
6= mean magnitudes derived in the present paper from data in \citet{Berdnikov2008} (for V1359 Aql, BC Aql, FQ Lac, QQ Per, SU Sct, and V526 Aql),
the {\it Optical Monitor Camera} (OMC) on board {\it INTEGRAL} \citep{OMC2003}\footnote{\url{https://sdc.cab.inta-csic.es/omc/secure/form\_busqueda.jsp}}
(V1048 Cen, V556 Cas, V1019 Cas, and V411 Lac), and \citet{GarciaM01} for LO Cam.
7= APASS (AAVSO Photometric All Sky Survey, \citealt{Henden16}) with correction of $-0.03$ mag applied.
Column~6-9. $JHK$-band magnitude and reference: 
1= \citet{MP11}, 2= \citet{Laney1992}, 3= Laney (priv. comm.), as quoted in \citet{Genovali2014}, 4= \citet{Feast2008},  
5= \citet{Barnes1997}, 6= \citet{Welch1984},  
7= \citet{Genovali2014}, based on single-epoch 2MASS data and template fitting, 
8= \citet{Schechter1992}, 9= \citet{McGonegal83}, 
10= 2MASS, 11= DENIS, 12= UKIDSS DR6,
13= $\alpha$ UMi: COBE-DIRBE flux at 1.25 and 2.2 $\mu$m converted to 2MASS, see main text. 
Photometry from (1) - (6) are intensity-mean magnitudes and are in the original photometric system.
If multiple single-epoch data was used to obtain the median magnitude, all references used are listed and the available 
number of $K$ magnitudes is given after the `=' sign.
In that case, the available magnitudes were de-reddened, then transformed to the 2MASS system, and reddened again. 
The magnitudes quoted are thus `observed' magnitudes on the 2MASS system.  
In the case of single-epoch 2MASS data the quality flag is listed if it is not AAA.
Columns~10-11. Reddening value with error, and Reference. Some corrections have been applied, see main text. 
1= Fernie et al. (1995)\footnote{\url{http://www.astro.utoronto.ca/DDO/research/cepheids/table_colourexcess.html}}, multiplied by 0.94,
2= \citet{Acharova2012} without scaling,
3= \citet{LL11} (scaling by 0.99), 
4= \citet{CC87} (scaling by 0.987), 
5= \citet{Kashuba16} (scaling by 0.94),
6= \citet{MarAnd15} (scaling by 0.97), 
7= \citet{Sziladi07} (scaling by 0.92),
8= based on 3D reddening models (see text),
9= \citet{Andrievsky2016} without scaling.
Columns~12, 14, 16. Three different [Fe/H] values, and References (Cols.~ 13, 15, 17).
1= \citet{Genovali2014}, 2= \citet{Genovali2015}, 
3= \citet{Ngeow12}, 4= \citet{LL11},
5= \citet{Acharova2012}, with correction +0.055 dex applied, 
6= \citet{Sziladi07}, with correction +0.032 dex applied,
7= \citet{MarAnd15}, with correction +0.030 dex applied.
}
\end{landscape}
}

\setcounter{table}{0}
\longtab[1]{
\def\thetable{A.\arabic{table}}
\begin{longtable}{ccrl}
\caption{\label{Tab:App1} $PL$ relations of the form $M = \alpha + \beta \log P$; 
$N$ is the number of stars in the solution. } \\
\hline  \hline
   $\alpha$   & $\beta$    & $N$ & Constraints  \\  
\hline
\endfirsthead
\caption{continued.}\\
\hline\hline
   $\alpha$   & $\beta$    & $N$ & Constraints  \\  
\hline
\endhead
\hline
\endfoot
 \multicolumn{4}{c}{ \underline {GDR2 parallaxes} } \\
 \multicolumn{4}{c}{ \em $V$ band } \\

 -1.984 $\pm$ 0.121 & -2.326 $\pm$ 0.115 & 425 & all                     \\  
 -1.985 $\pm$ 0.136 & -2.321 $\pm$ 0.134 & 419 & $\mid$GOF$\mid<100$     \\  
 -1.956 $\pm$ 0.043 & -2.337 $\pm$ 0.049 & 376 & $\mid$GOF$\mid<20$      \\  
 -2.033 $\pm$ 0.051 & -2.220 $\pm$ 0.061 & 293 & $\mid$GOF$\mid<10$      \\  
 -2.049 $\pm$ 0.055 & -2.201 $\pm$ 0.066 & 250 & $\mid$GOF$\mid<8$       \\  
 -2.138 $\pm$ 0.074 & -2.084 $\pm$ 0.092 & 197 & $\mid$GOF$\mid<6$       \\  
 -2.134 $\pm$ 0.090 & -2.093 $\pm$ 0.110 & 113 & $\mid$GOF$\mid<4$       \\  

 -2.047 $\pm$ 0.055 & -2.202 $\pm$ 0.067 & 248 & $\mid$GOF$\mid<8$, $\epsilon_i<0.001$  \\  

\\
 & & &  $\mid$GOF$\mid<8$, $\epsilon_i<0.001$, and ... \\
 -1.881 $\pm$ 0.057 & -2.416 $\pm$ 0.065 & 211 & FU                   \\  %
 -2.661 $\pm$ 0.176 & -1.344 $\pm$ 0.247 &  37 & FO                   \\  %
 -1.852 $\pm$ 0.060 & -2.446 $\pm$ 0.068 & 205 & FU, $P >$2.5d         \\  %
 -1.857 $\pm$ 0.060 & -2.441 $\pm$ 0.068 & 200 & FU, $P >$3.0d         \\  %
 -1.869 $\pm$ 0.059 & -2.431 $\pm$ 0.070 & 185 & FU, $P >$3.5d         \\  %

 -1.911 $\pm$ 0.080 & -2.393 $\pm$ 0.093 & 159 & FU, $P >$2.5d, $\mid$GOF$\mid<6$  \\  
 -1.882 $\pm$ 0.115 & -2.363 $\pm$ 0.131 &  90 & FU, $P >$2.5d, $\mid$GOF$\mid<4$  \\  

 -1.847 $\pm$ 0.072 & -2.457 $\pm$ 0.086 & 202 & FU, $P >$2.5d, $P<$35      \\  

 -1.886 $\pm$ 0.060 & -2.415 $\pm$ 0.068 & 207 & FU, no uncertain types        \\  

 -1.794 $\pm$ 0.058 & -2.385 $\pm$ 0.066 & 205 & FU, $P >$2.5d, ZPoff= $-0.029$ mas    \\  
 -1.759 $\pm$ 0.057 & -2.353 $\pm$ 0.065 & 205 & FU, $P >$2.5d, ZPoff= $-0.046$ mas    \\  

 -1.883 $\pm$ 0.113 & -2.443 $\pm$ 0.126 & 205 & FU, $P >$2.5d, $\delta_{\rm PL}= 0.20$ \\  

\\
 \multicolumn{4}{c}{ \em $K$ band } \\

 -2.980 $\pm$ 0.031 & -3.060 $\pm$ 0.038 & 425 & all                      \\  
 -2.985 $\pm$ 0.029 & -3.051 $\pm$ 0.035 & 419 & $\mid$GOF$\mid<100$      \\  
 -3.008 $\pm$ 0.031 & -3.023 $\pm$ 0.036 & 376 & $\mid$GOF$\mid<20$      \\  
 -3.073 $\pm$ 0.036 & -2.929 $\pm$ 0.045 & 294 & $\mid$GOF$\mid<10$      \\  
 -3.083 $\pm$ 0.039 & -2.907 $\pm$ 0.048 & 250 & $\mid$GOF$\mid<8$      \\  
 -3.171 $\pm$ 0.053 & -2.787 $\pm$ 0.067 & 197 & $\mid$GOF$\mid<6$      \\  
 -3.209 $\pm$ 0.063 & -2.769 $\pm$ 0.081 & 113 & $\mid$GOF$\mid<4$      \\  

 -3.081 $\pm$ 0.039 & -2.907 $\pm$ 0.045 & 248 & $\mid$GOF$\mid<8$, $\epsilon_i<0.001$     \\  

\\
 & & &  $\mid$GOF$\mid<8$, $\epsilon_i<0.001$, and ... \\
 -2.985 $\pm$ 0.041 & -3.083 $\pm$ 0.048 & 211 & FU                   \\  
 -3.622 $\pm$ 0.125 & -1.974 $\pm$ 0.192 &  37 & FO                   \\  
 -2.953 $\pm$ 0.044 & -3.119 $\pm$ 0.051 & 205 & FU, $P >$2.5d        \\  
 -2.964 $\pm$ 0.044 & -3.107 $\pm$ 0.052 & 200 & FU, $P >$3.0d        \\  
 -2.992 $\pm$ 0.052 & -3.079 $\pm$ 0.060 & 185 & FU, $P >$3.5d        \\  

 -3.017 $\pm$ 0.057 & -3.053 $\pm$ 0.069 & 159 & FU, $P >$2.5d, $\mid$GOF$\mid<6$  \\  
 -3.146 $\pm$ 0.075 & -2.869 $\pm$ 0.094 &  90 & FU, $P >$2.5d, $\mid$GOF$\mid<4$  \\  
 -3.181 $\pm$ 0.091 & -2.829 $\pm$ 0.107 &  76 & FU, $P >$3.5d, $\mid$GOF$\mid<4$  \\  

 -2.957 $\pm$ 0.044 & -3.117 $\pm$ 0.053 & 201 & FU, $P >$2.5d, no uncertain        \\  
 -2.918 $\pm$ 0.047 & -3.166 $\pm$ 0.060 & 202 & FU, $P >$2.5d, $P<$35              \\  

 -2.894 $\pm$ 0.042 & -3.055 $\pm$ 0.048 & 205 & FU, $P >$2.5d, ZPoff= $-0.029$ mas      \\  
 -2.861 $\pm$ 0.040 & -3.021 $\pm$ 0.047 & 205 & FU, $P >$2.5d, ZPoff= $-0.046$ mas      \\  
 -2.957 $\pm$ 0.053 & -3.114 $\pm$ 0.061 & 205 & FU, $P >$2.5d, $\delta_{\rm PL}= 0.066$     \\  

\\
 \multicolumn{4}{c}{ \em $WVK$ band } \\

 -3.114 $\pm$ 0.031 & -3.158 $\pm$ 0.038 & 425 & all                    \\  
 -3.118 $\pm$ 0.030 & -3.149 $\pm$ 0.036 & 419 & $\mid$GOF$\mid<100$    \\  
 -3.149 $\pm$ 0.031 & -3.113 $\pm$ 0.037 & 376 & $\mid$GOF$\mid<20$    \\  
 -3.213 $\pm$ 0.035 & -3.020 $\pm$ 0.045 & 293 & $\mid$GOF$\mid<10$    \\  
 -3.221 $\pm$ 0.038 & -2.998 $\pm$ 0.048 & 250 & $\mid$GOF$\mid<8$    \\  
 -3.311 $\pm$ 0.052 & -2.876 $\pm$ 0.068 & 197 & $\mid$GOF$\mid<6$    \\  
 -3.360 $\pm$ 0.063 & -2.849 $\pm$ 0.082 & 113 & $\mid$GOF$\mid<4$    \\  

 -3.221 $\pm$ 0.038 & -2.998 $\pm$ 0.045 & 248 & $\mid$GOF$\mid<8$, $\epsilon_i<0.001$   \\  

\\
& & &  $\mid$GOF$\mid<8$, $\epsilon_i<0.001$, and ... \\
 -3.136 $\pm$ 0.042 & -3.167 $\pm$ 0.049 & 211 & FU                   \\  
 -3.747 $\pm$ 0.134 & -2.027 $\pm$ 0.204 &  37 & FO                   \\  
 -3.101 $\pm$ 0.043 & -3.203 $\pm$ 0.051 & 205 & FU, $P >$2.5d         \\  
 -3.113 $\pm$ 0.043 & -3.191 $\pm$ 0.052 & 200 & FU, $P >$3.0d         \\  
 -3.145 $\pm$ 0.051 & -3.159 $\pm$ 0.060 & 200 & FU, $P >$3.5d         \\  

 -3.168 $\pm$ 0.057 & -3.133 $\pm$ 0.068 & 159 & FU, $P >$2.5d, $\mid$GOF$\mid<6$   \\  
 -3.322 $\pm$ 0.075 & -2.926 $\pm$ 0.092 &  90 & FU, $P >$2.5d, $\mid$GOF$\mid<4$   \\  
 -3.374 $\pm$ 0.091 & -2.864 $\pm$ 0.106 &  76 & FU, $P >$3.5d, $\mid$GOF$\mid<4$   \\  

 -3.105 $\pm$ 0.044 & -3.202 $\pm$ 0.053 & 201 & FU, $P >$2.5d, no uncertain   \\  
 -3.062 $\pm$ 0.048 & -3.256 $\pm$ 0.060 & 202 & FU, $P >$2.5d, $P<$35    \\  

 -3.043 $\pm$ 0.041 & -3.140 $\pm$ 0.048 & 205 & FU, $P >$2.5d, ZPoff= $-0.029$ mas   \\  
 -3.009 $\pm$ 0.040 & -3.105 $\pm$ 0.046 & 205 & FU, $P >$2.5d, ZPoff= $-0.046$ mas   \\  
 -3.103 $\pm$ 0.049 & -3.201 $\pm$ 0.057 & 205 & FU, $P >$2.5d, $\delta_{\rm PL}= 0.049$   \\  

 -3.064 $\pm$ 0.054 & -3.257 $\pm$ 0.066 & 202 & FU, $P >$2.5d, $\delta_{\rm PL}= 0.049$, $P<$35  \\  
 -3.032 $\pm$ 0.058 & -3.278 $\pm$ 0.070 & 157 & FU, $P >$2.5d, $\delta_{\rm PL}= 0.049$, $P<$35, $\pi >0, \spi/\pi<0.2$  \\  
 -2.994 $\pm$ 0.065 & -3.296 $\pm$ 0.082 &  85 & FU, $P >$2.5d, $\delta_{\rm PL}= 0.049$, $P<$35, $\pi >0, \spi/\pi<0.1$  \\  
 -2.995 $\pm$ 0.118 & -3.294 $\pm$ 0.159 &  23 & FU, $P >$2.5d, $\delta_{\rm PL}= 0.049$, $P<$35, $\pi >0, \spi/\pi<0.05$  \\  

\\
 \multicolumn{4}{c}{ \underline {GDR2+external parallaxes} } \\
 \multicolumn{4}{c}{ \em $V$ band } \\

 -1.924 $\pm$ 0.046 & -2.355 $\pm$ 0.054 & 426 & all             \\  
 -1.931 $\pm$ 0.045 & -2.353 $\pm$ 0.054 & 425 & no Polaris     \\
 -1.813 $\pm$ 0.051 & -2.490 $\pm$ 0.057 & 374 & FU              \\ 
 -2.417 $\pm$ 0.144 & -1.680 $\pm$ 0.197 &  52 & FO              \\ 
 -2.437 $\pm$ 0.139 & -1.659 $\pm$ 0.196 &  51 & FO, no Polaris    \\ 

 -1.813 $\pm$ 0.052 & -2.490 $\pm$ 0.059 & 373 & FU, GOF $<100$  \\
 -1.831 $\pm$ 0.045 & -2.464 $\pm$ 0.051 & 345 & FU, GOF $<20$  \\ 
 -1.887 $\pm$ 0.051 & -2.379 $\pm$ 0.058 & 265 & FU, GOF $<10$  \\ 
 -1.882 $\pm$ 0.058 & -2.380 $\pm$ 0.065 & 225 & FU, GOF $<8$   \\ 

 -1.878 $\pm$ 0.054 & -2.384 $\pm$ 0.062 & 223 & FU, GOF $<8$, $\epsilon_i<0.001$  \\ 

\\
 & & & $\mid$GOF$\mid<8$, $\epsilon_i<0.001$, and ...  \\

 -1.856 $\pm$ 0.058 & -2.407 $\pm$ 0.068 & 217 & FU, $P >$2.5d  \\ 
 -1.856 $\pm$ 0.060 & -2.405 $\pm$ 0.066 & 212 & FU, $P >$3.0d  \\ 
 -1.852 $\pm$ 0.055 & -2.411 $\pm$ 0.064 & 197 & FU, $P >$3.5d  \\ 

 -1.858 $\pm$ 0.056 & -2.410 $\pm$ 0.065 & 213 & FU, $P >$2.5d, no uncertain  \\ 
 -1.840 $\pm$ 0.070 & -2.430 $\pm$ 0.081 & 213 & FU, $P >$2.5d, $P<$35   \\ 

 -1.795 $\pm$ 0.057 & -2.361 $\pm$ 0.066 & 217 & FU, $P >$2.5d, ZPoff= $-0.029$ mas   \\ 
 -1.760 $\pm$ 0.056 & -2.335 $\pm$ 0.065 & 217 & FU, $P >$2.5d, ZPoff= $-0.046$ mas   \\ 

 -1.882 $\pm$ 0.105 & -2.406 $\pm$ 0.117 & 217 & FU, $P >$2.5d, $\delta_{\rm PL}= 0.20$  \\ 

 -1.886 $\pm$ 0.056 & -2.378 $\pm$ 0.063 & 221 & FU, GOF $<8$, no uncertain   \\ 
 -1.812 $\pm$ 0.054 & -2.346 $\pm$ 0.061 & 221 & FU, GOF $<8$, no uncertain, ZPoff= $-0.029$ mas   \\
 -1.771 $\pm$ 0.054 & -2.328 $\pm$ 0.062 & 221 & FU, GOF $<8$, no uncertain, ZPoff= $-0.046$ mas   \\
 -1.767 $\pm$ 0.062 & -2.323 $\pm$ 0.069 & 221 & FU, GOF $<8$, no uncertain, ZPoff= $-0.046$ mas, $\spi \cdot 1.3$  \\

 -2.14  $\pm$ 0.18  & -1.98  $\pm$ 0.27  & 112 & FU, GOF $<8$, $P < 6$  \\
 -1.81  $\pm$ 0.085 & -2.45  $\pm$ 0.085 & 113 & FU, GOF $<8$, $P > 6$  \\

 -1.87  $\pm$ 0.10  & -2.40  $\pm$ 0.14  & 157 & FU, GOF $<8$, $P < 8$  \\
 -1.68  $\pm$ 0.15  & -2.54  $\pm$ 0.13  &  68 & FU, GOF $<8$, $P > 8$  \\

 -1.870 $\pm$ 0.060 & -2.433 $\pm$ 0.069 & 358 & FU, $V > 6.3$  \\
 -1.884 $\pm$ 0.059 & -2.411 $\pm$ 0.068 & 214 & FU, $V > 6.3$, GOF $<8$  \\
 -1.539 $\pm$ 0.035 & -2.81 fixed  & 1   & Polaris \\
 -1.542 $\pm$ 0.015 & -2.81 fixed  & 225 & FU, GOF $<8$   \\

\\
 \multicolumn{4}{c}{ \em $K$ band } \\
 -2.912 $\pm$ 0.029 & -3.076 $\pm$ 0.035 & 426 & all             \\ 
 -2.907 $\pm$ 0.034 & -3.151 $\pm$ 0.040 & 374 & FU              \\ 
 -2.977 $\pm$ 0.042 & -3.048 $\pm$ 0.049 & 225 & FU, GOF $<8$             \\ 
 -2.981 $\pm$ 0.039 & -3.045 $\pm$ 0.047 & 221 & FU, GOF $<8$, no uncertain   \\ 

\\
& & & $\mid$GOF$\mid<8$, $\epsilon_i<0.001$, and ... \\

 -2.943 $\pm$ 0.042 & -3.084 $\pm$ 0.051 & 217 & FU, GOF $<8$, $P > 2.5$d   \\ 
 -2.950 $\pm$ 0.042 & -3.075 $\pm$ 0.048 & 212 & FU, GOF $<8$, $P > 3.0$d   \\ 
 -2.957 $\pm$ 0.049 & -3.069 $\pm$ 0.056 & 197 & FU, GOF $<8$, $P > 3.5$d   \\ 

 -2.911 $\pm$ 0.048 & -3.130 $\pm$ 0.058 & 213 & FU, GOF $<8$, $P > 2.5$, $P<$35d    \\ 

 -2.882 $\pm$ 0.040 & -3.037 $\pm$ 0.048 & 217 & FU, GOF $<8$, $P > 2.5$d, ZPoff= $-0.029$ mas     \\ 
 -2.848 $\pm$ 0.039 & -3.011 $\pm$ 0.047 & 217 & FU, GOF $<8$, $P > 2.5$d, ZPoff= $-0.046$ mas     \\ 

 -2.946 $\pm$ 0.052 & -3.084 $\pm$ 0.060 & 217 & FU, GOF $<8$, $P > 2.5$d, $\delta_{\rm PL}= 0.066$  \\ 

 -2.976 $\pm$ 0.042 & -3.051 $\pm$ 0.050 & 220 & FU, GOF $<8$, no uncertain, $P > 2$d   \\
 -2.956 $\pm$ 0.042 & -3.074 $\pm$ 0.051 & 209 & FU, GOF $<8$, no uncertain, $P > 3$d   \\
 -2.959 $\pm$ 0.050 & -3.070 $\pm$ 0.058 & 180 & FU, GOF $<8$, no uncertain, $P > 4$d   \\
 -2.805 $\pm$ 0.061 & -3.214 $\pm$ 0.066 & 140 & FU, GOF $<8$, no uncertain, $P > 5$d   \\
 -2.915 $\pm$ 0.076 & -3.119 $\pm$ 0.077 & 110 & FU, GOF $<8$, no uncertain, $P > 6$d   \\
 -2.917 $\pm$ 0.071 & -3.012 $\pm$ 0.072 & 110 & FU, GOF $<8$, no uncertain, $P > 6$d, ZPoff= $-0.029$ mas   \\
 -2.915 $\pm$ 0.071 & -2.956 $\pm$ 0.070 & 110 & FU, GOF $<8$, no uncertain, $P > 6$d, ZPoff= $-0.046$ mas   \\
 -2.897 $\pm$ 0.095 & -3.111 $\pm$ 0.096 & 110 & FU, GOF $<8$, no uncertain, $P > 6$d, ZPoff= $-0.000$ mas, $\spi \cdot 1.3$   \\

 -2.765 $\pm$ 0.008 & -3.26 fixed & 426 & all             \\
 -2.484 $\pm$ 0.028 & -3.26 fixed &   1 & Polaris          \\
 -2.735 $\pm$ 0.010 & -3.26 fixed & 269 & GOF $<8$         \\ 
 -2.807 $\pm$ 0.010 & -3.26 fixed & 225 & FU, GOF $<8$      \\ 

\\
 \multicolumn{4}{c}{ \em $WVK$ band } \\
 -3.045 $\pm$ 0.030 & -3.174 $\pm$ 0.035 & 426 & all             \\  
 -3.122 $\pm$ 0.036 & -3.026 $\pm$ 0.045 & 269 & GOF $<8$        \\ 
 -3.125 $\pm$ 0.042 & -3.133 $\pm$ 0.048 & 225 & GOF $<8$, FU     \\
 -3.609 $\pm$ 0.042 & -2.093 $\pm$ 0.200 &  44 & GOF $<8$, FO     \\

 -3.092 $\pm$ 0.042 & -3.166 $\pm$ 0.050 & 218 & GOF $<8$, FU, $P > 2.5$d   \\
 -3.099 $\pm$ 0.041 & -3.161 $\pm$ 0.047 & 213 & GOF $<8$, FU, $P > 3.0$d   \\
 -3.108 $\pm$ 0.048 & -3.151 $\pm$ 0.055 & 197 & GOF $<8$, FU, $P > 3.5$d   \\

 -3.060 $\pm$ 0.048 & -3.210 $\pm$ 0.060 & 214 & GOF $<8$, FU, $P > 2.5$, $P<$35d   \\

 -3.031 $\pm$ 0.041 & -3.119 $\pm$ 0.048 & 218 & GOF $<8$, FU, $P > 2.5$d, ZPoff= $-0.029$ mas   \\
 -2.996 $\pm$ 0.040 & -3.093 $\pm$ 0.048 & 218 & GOF $<8$, FU, $P > 2.5$d, ZPoff= $-0.046$ mas   \\

 -3.095 $\pm$ 0.047 & -3.166 $\pm$ 0.053 & 218 & GOF $<8$, FU, $P > 2.5$d, $\delta_{\rm PL}= 0.049$   \\

\end{longtable}
}

\end{document}